\documentclass[twocolumn,amsmath,amssymb,superscriptaddress]{revtex4-1}

\usepackage{graphicx}

\usepackage{amstext,amsmath,amssymb,amsfonts}
\usepackage{bm}

\usepackage{txfonts}
\usepackage{mathtools}
\DeclareMathAlphabet{\mathpzc}{OT1}{pzc}{m}{it}
\DeclareFontFamily{OT1}{pzc}{}
\DeclareFontShape{OT1}{pzc}{m}{it}{<-> s * [1.100] pzcmi7t}{}
\DeclareMathAlphabet{\mathpzc}{OT1}{pzc}{m}{it}

\usepackage{hyperref}

\usepackage{color}
\definecolor{lightblue}{rgb}{0.2,0.2,0.7}
\definecolor{darkblue}{rgb}{0,0.25,0.5}
\definecolor{redbrown}{rgb}{0.875,0.25,0.125}
\definecolor{darkgreen}{rgb}{0,0.5,0}

\newcommand{\bra}[1]{\ensuremath{\langle #1 \vert}}
\newcommand{\ket}[1]{\ensuremath{\vert #1  \rangle}}
\newcommand{\braket}[2]{\ensuremath{\langle  #1 \vert #2  \rangle}}
\renewcommand{\b}[1]{\ensuremath{\mathbf{#1}}}
\renewcommand{\H}{\ensuremath{\text{H}}}

\renewcommand{\l}{\ensuremath{\lambda}}

\newcommand{\Tr}{\ensuremath{\text{Tr}}}
\newcommand{\tr}{\ensuremath{\text{tr}}}

\newcommand{\HF}{\ensuremath{\text{HF}}}

\renewcommand{\S}{\ensuremath{\text{S}}}

\renewcommand{\d}{\ensuremath{\text{d}}}

\newcommand{\g}{\ensuremath{\text{g}}}
\renewcommand{\u}{\ensuremath{\text{u}}}
\newcommand{\x}{\ensuremath{\text{x}}}

\renewcommand{\c}{\ensuremath{\text{c}}}

\DeclareMathOperator{\sgn}{sgn}

\renewcommand{\L}{\ensuremath{\text{L}}}

\renewcommand{\i}{\ensuremath{\text{i}}}

\newcommand{\calh}{\ensuremath{\mathpzc{h}}}

\newcommand{\p}{\ensuremath{\text{p}}}
\renewcommand{\g}{\ensuremath{\text{g}}}
\renewcommand{\u}{\ensuremath{\text{u}}}
\newcommand{\PS}{\ensuremath{\text{PS}}}
\newcommand{\NS}{\ensuremath{\text{NS}}}
\newcommand{\vp}{\ensuremath{\text{vp}}}
\newcommand{\el}{\ensuremath{\text{el}}}

\begin{document}

\title{Effective quantum electrodynamics: One-dimensional model of the relativistic hydrogen-like atom}
\author{Timoth\'ee Audinet}
\email{timothee.audinet@sorbonne-universite.fr}
\affiliation{Laboratoire de Chimie Th\'eorique, Sorbonne Universit\'e and CNRS, F-75005 Paris, France}
\author{Julien Toulouse}
\email{toulouse@lct.jussieu.fr}
\affiliation{Laboratoire de Chimie Th\'eorique, Sorbonne Universit\'e and CNRS, F-75005 Paris, France}
\affiliation{Institut Universitaire de France, F-75005 Paris, France}

\date{May 22, 2023}

\begin{abstract}
We consider a one-dimensional effective quantum electrodynamics (QED) model of the relativistic hydrogen-like atom using delta-potential interactions. We discuss the general exact theory and the Hartree-Fock approximation. The present one-dimensional effective QED model shares the essential physical feature of the three-dimensional theory: the nuclear charge polarizes the vacuum state (creation of electron-positron pairs) which results in a QED Lamb-type shift of the bound-state energy. Yet, this 1D effective QED model eliminates some of the most serious technical difficulties of the three-dimensional theory coming from renormalization. We show how to calculate the vacuum-polarization density at zeroth order in the two-particle interaction and the QED Lamb-type shift of the bound-state energy at first order in the two-particle interaction. The present work may be considered as a step toward the development of a quantum-chemistry effective QED theory of atoms and molecules.
\end{abstract}

\maketitle

\section{Introduction}

It is important to take into account the effects of special relativity in the quantum description of chemical systems with heavy elements~\cite{Pyy-ARPC-12}. Relativistic electronic-structure computational methods based on the no-pair Dirac-Coulomb or Dirac-Coulomb-Breit Hamiltonian have thus been developed and are now routinely applied on molecular systems (see, e.g., Refs.~\onlinecite{SauVis-INC-03,DyaFae-BOOK-07,ReiWol-BOOK-09}). The next challenge for relativistic quantum chemistry is to go beyond the no-pair approximation~\cite{Suc-PRA-80,Mit-PRA-81}, i.e. including the quantum-electrodynamics (QED) effect of virtual electron-positron pairs. This is desirable not only for an increased accuracy but also in order to put relativistic quantum chemistry on deeper theoretical grounds.

Bound-state QED perturbative techniques have been developed to perform highly accurate calculations on few-electron atomic systems (see, e.g., Refs.~\onlinecite{MohPluSof-PR-98,Sha-PR-02,LinSalAse-PR-04,IndMoh-INC-16}). For many-electron atoms, it has been proposed to estimate QED corrections with model one-electron operators (see, e.g., Refs.~\onlinecite{PyyZha-JPB-03,ShaTupYer-PRA-13,SchPasPunBow-NPA-15,PasEliBorKalSch-PRL-17,MalGlaShaTupYerZay-PRA-22,Sal-THESIS-22}). This approach has also been extended to many-electron molecular systems (see, e.g., Refs.~\onlinecite{Skr-JCP-21,SunSalSau-JCP-22}). Another strategy to include QED effects in electronic-structure calculations of atoms and molecules would be to use relativistic density-functional theory based on QED~\cite{DreGro-BOOK-90,EngMulSpeDre-INC-95,EngDre-INC-96,Eng-INC-02,EngDre-BOOK-11,EscSer-JCC-99,Esc-BOOK-03} but it has yet to be applied beyond the no-pair approximation.

An attractive approach to perform ab initio calculations beyond the no-pair approximation is to use a fermionic Fock-space effective QED Hamiltonian with the Coulomb or Coulomb-Breit two-particle interaction (see, e.g., Refs.~\cite{ChaIra-JPB-89,SauVis-INC-03,Kut-CP-12,LiuLin-JCP-13,Liu-PR-14,Liu-IJQC-15,Liu-JCP-20,Tou-SPC-21,Liu-WIRES-22}). This effective QED theory properly includes the effects of vacuum polarization through the creation of electron-positron pairs but does not include explicitly the photon degrees of freedom. It is thus a more tractable alternative to full QED for atomic and molecular calculations. This so-called no-photon QED has been the subject of a number of detailed mathematical studies~\cite{HaiLewSer-CMP-05,HaiLewSer-JPA-05,HaiLewSol-CPAM-07,HaiLewSerSol-PRA-07,GraLewSer-CMP-09,HaiLewSer-ARMA-09,GraLewSer-CMP-11,Lew-INC-11}, which in particular established the soundness of this approach at the Hartree-Fock level. Based on this effective QED theory, it has been proposed to formulate a relativistic density-functional theory~\cite{Tou-SPC-21} and a relativistic reduced density-matrix functional theory~\cite{RodGieVis-SPC-22}. However, as in full QED, this effective QED theory still contains ultraviolet divergences that need to be dealt with by regularization and renormalization (see, e.g., Refs.~\onlinecite{HaiLewSerSol-PRA-07,Lew-INC-11}). Consequently, no practical implementation of this effective QED theory has been done so far.

In this work, as a first step toward the implementation of the above-mentioned effective QED theory for atomic and molecular calculations, we apply it to a one-dimensional (1D) model of the relativistic hydrogen-like atom using delta-potential interactions. In the non-relativistic version of this model~\cite{Fro-JCP-56,HerSti-PRA-75,Her-JCP-86,TraGinTou-JCP-22}, the use of the delta potential is motivated by the fact that it leads to the same ground-state energy and wave function as the ground-state energy and radial wave function of the three-dimensional (3D) hydrogen-like atom with the Coulomb potential. The relativistic version of this model has also been studied without QED effects~\cite{SubBha-JPC-72,Lap-AJP-83,FilLorBan-JPA-12,GuiMunPirSan-FP-19} and some QED aspects were considered by Nogami and Beachey~\cite{NogBea-EL-86}. The present 1D effective QED model can also be thought of as the massive Thirring quantum-field-theory model (see, e.g., Refs.~\onlinecite{Thi-AP-58,BerTha-PRL-79,BerTha-PRD-79,Tha-RMP-81,AbdAbdRot-BOOK-01}) with an additional external potential. Note that the massive Thirring model can itself be essentially thought of as a sort of ``infinite-mass photon'' limit~\cite{DubTar-AP-67,Kon-PTP-97,BufCasPim-IJMPA-11} of the massive Thirring-Wess model~\cite{ThiWes-AP-64} (i.e., the massive Schwinger model of QED~\cite{Sch-PR-62a} with a massive photon), the infinite-mass photon field generating the delta-potential interaction. As we will show, the present 1D effective QED model shares the essential physical feature of the 3D effective QED theory that we are interested in: the nuclear charge polarizes the vacuum state (creation of electron-positron pairs) which results in a Lamb-type shift of the bound-state energy. Yet, this 1D effective QED model evacuates a lot of difficulties of the 3D effective QED theory: it removes some of the most serious ultraviolet divergences that appear in the standard 3D case. 

The paper is organized as follows. In Section~\ref{sec:firstquantized}, we consider the first-quantized theory of a 1D relativistic electron in free space and in the hydrogen-like atom with a delta potential. In Section~\ref{sec:secondquantized}, we formulate the second-quantized effective QED theory for the 1D hydrogen-like atom. After writing the general Hartree-Fock equations, we study the vacuum-polarization density and the Lamb-type shift of the bound-state energy in a first-order perturbation theory with respect to the Coulomb-Breit-type two-particle interaction. Section~\ref{sec:conclusion} contains our conclusions. Finally, some technical details are given in Appendices~\ref{app:3dto1d}-\ref{app:Nvac}. Hartree atomic units (a.u.) are used throughout this work.

\section{First-quantized one-electron theory}
\label{sec:firstquantized}

In this section, we consider a 1D relativistic electron in a first-quantized theory. As shown in Appendix~\ref{app:3dto1d}, in 1D we can work with two-component states in the Hilbert space $\calh  = L^2(\mathbb{R},\mathbb{C}) \otimes \mathbb{C}^2$.

\subsection{Free-electron Dirac equation}

Let us start with the 1D free-electron Dirac equation (see, e.g., Refs.~\onlinecite{GuiMunPirSan-FP-19,KarIshPos-TCA-20})
\begin{eqnarray}
\b{D}_0(x) \bm{\psi}(x) = \varepsilon \bm{\psi}(x),
\label{Dpsi=epspsi}
\end{eqnarray}
where $\bm{\psi}(x)$ is a two-component vector with large (L) and small (S) components
\begin{eqnarray}
\bm{\psi}(x)= \left(\begin{array}{c} \psi^\L(x)\\\psi^\S(x)\end{array}\right), 
\end{eqnarray}
$\varepsilon$ is the associated energy, and $\b{D}_0$ is the 1D free-electron $2\times2$ Dirac Hamiltonian (see Appendix~\ref{app:3dto1d})
\begin{eqnarray}
\b{D}_0(x) = c \bm{\sigma}_1 \; p_x + \bm{\sigma}_3 \; mc^2,
\label{freeDiracHam}
\end{eqnarray}
where $p_x= -\i \d/\d x$ is the momentum operator, $c$ is the speed of light, $m = 1$ a.u. is the electron mass (which will be kept in the equations for clarity), and $\bm{\sigma}_1$ and $\bm{\sigma}_3$ are the $2\times2$ Pauli matrices
\begin{eqnarray}
\bm{\sigma}_1 = \left(\begin{array}{cc}
0&1\\
1&0\\
\end{array}\right)
~~\text{and}~~
\bm{\sigma}_3 = \left(\begin{array}{cc}
1&0\\
0&-1\\
\end{array}\right).
\end{eqnarray}
The domain of this Hamiltonian (i.e., the set of functions on which it can act) is $\text{Dom}(\b{D}_0) = H^1(\mathbb{R},\mathbb{C}) \otimes \mathbb{C}^2$ where $H^1(\mathbb{R},\mathbb{C})=\{\psi\in L^2(\mathbb{R},\mathbb{C}) \;|\; \d\psi/\d x \in L^2(\mathbb{R},\mathbb{C})\}$ is the first-order Sobolev space.

The Hamiltonian $\b{D}_0$ has parity symmetry, i.e. it commutes with the relativistic $2\times2$ parity operator
\begin{eqnarray}
\b{P} = \bm{\sigma}_3 P_{x \to -x},
\end{eqnarray}
where $P_{x \to -x}$ is the spatial parity operator which flips the sign of the coordinate $x$. We can thus look for gerade (g) and ungerade (u) symmetry-adapted eigenfunctions of $\b{D}_0$ such that 
$\b{P} \bm{\psi}^\g(x) = \bm{\psi}^\g(x)$ and $\b{P} \bm{\psi}^\u(x) = -\bm{\psi}^\u(x)$.

As well known, the Hamiltonian $\b{D}_0$ has only a continuous energy spectrum $(-\infty,-mc^2]\cup [mc^2,+\infty)$. The generalized eigenfunctions (i.e., ``continuum eigenfunctions'' not belonging to the Hilbert space ${\calh}$) associated with the positive energy $\varepsilon_k = \sqrt{k^2 c^2 + m^2 c^4}$ are
\begin{subequations}
\begin{gather}
\bm{\psi}_{+,k}^{\g} (x)= A_k \left(\!\begin{array}{c}
\cos(kx)\\
\i s_k \sin(kx)
\end{array}\!\right), \; k \in [0,+\infty),\\
\bm{\psi}_{+,k}^{\u} (x)= A_k \left(\!\begin{array}{c}
\sin(kx)\\
-\i s_k \cos(kx)
\end{array}\!\right), \; k \in (0,+\infty),
\end{gather}
\label{electronicFreeWF}
\end{subequations}
and the generalized eigenfunctions associated with the negative energy $-\varepsilon_k$ are
\begin{subequations}
\begin{gather}
\bm{\psi}_{-,k}^{\g} (x)= A_k \left(\!\begin{array}{c}
\i s_k \cos(kx)\\
\sin(kx)
\end{array}\!\right), \; k \in (0,+\infty),\\
\bm{\psi}_{-,k}^{\u} (x)= A_k \left(\!\begin{array}{c}
-\i s_k \sin(kx)\\
\cos(kx)
\end{array}\!\right),\; k \in [0,+\infty),
\end{gather}
\label{positronicFreeWF}
\end{subequations}
where $s_k = k c /(\varepsilon_k + mc^2)$ and $A_k=\sqrt{(\varepsilon_k + mc^2)/(2\pi\varepsilon_k)}$ is a normalization constant chosen to impose the generalized orthogonality relation
\begin{eqnarray}
\int_{-\infty}^{\infty} \bm{\psi}_{\pm,k_1}^\dagger(x) \bm{\psi}_{\pm,k_2}^{}(x) \d x = \delta (k_1 - k_2).
\end{eqnarray}

Note that, in the non-relativistic limit ($c\to \infty$), we have $s_k \to 0$ and $A_k \to 1/\sqrt{\pi}$, and the generalized eigenfunctions properly reduce to the non-relativistic continuum states $(1/\sqrt{\pi}) \cos(kx)$ and $(1/\sqrt{\pi}) \sin(kx)$ (see, e.g., Ref.~\onlinecite{GriSch-BOOK-18}).

\subsection{Hydrogen-like Dirac equation}
\label{sec:hydrogenlike}

We now consider the 1D hydrogen-like Dirac equation~\cite{Lap-AJP-83}
\begin{eqnarray}
\b{D}(x) \tilde{\bm{\psi}}(x) = \tilde{\varepsilon} \tilde{\bm{\psi}}(x),
\label{DiraceqH}
\end{eqnarray}
with the 1D hydrogen-like Dirac Hamiltonian composed of the free-electron Dirac Hamiltonian and an electrostatic-type nuclear-electron Dirac-delta potential term
\begin{eqnarray}
\b{D}(x) = \b{D}_0(x) - Z \delta(x) \b{I}_2,
\label{Dv}
\end{eqnarray}
where $Z$ is the nuclear charge (with $0 \leq Z\leq 2c$ so as to have a positive bound-state energy in Eq.~\eqref{epsilon0tilde}) and $\b{I}_2$ is the $2\times2$ identity matrix. 

The delta potential in Eq.~(\ref{Dv}) is in fact ambiguous. Indeed, due to the delta potential, any eigenfunction $\tilde{\bm{\psi}}$ of $\b{D}$ is expected to have a discontinuity at $x=0$, but the action of a delta distribution on a discontinuous function is not a priori defined. Mathematically, $\b{D}$ can be precisely defined as a self-adjoint extension of the free-electron Dirac operator $\b{D}_0$ restricted to an initial domain of functions vanishing at $x=0$. This leads to defining $\b{D}$ as having the same action of $\b{D}_0$ but on a smaller domain of the form~\cite{Seb-LMP-89,BenDab-LMP-94,Hug-RMP-97,PanRic-JMP-14}
\begin{eqnarray}
\text{Dom}(\b{D}) = \left\{ \tilde{\bm{\psi}} \in H^1(\mathbb{R}\! \setminus \! \{0\},\mathbb{C}) \otimes \mathbb{C}^2 \; |\; \tilde{\bm{\psi}}(0^+) = \b{M} \tilde{\bm{\psi}}(0^-) \right\}, \;
\label{}
\end{eqnarray}
where $H^1(\mathbb{R}\! \setminus \! \{0\},\mathbb{C}) \equiv H^1(\mathbb{R}^{-},\mathbb{C})\oplus H^1(\mathbb{R}^{+},\mathbb{C})$ is a broken Sobolev space (i.e., the direct sum of Sobolev spaces on adjacent spatial domains without regularity conditions across the frontiers, here allowing for a discontinuity at $x=0$) and $\b{M}$ is a unitary $2\times 2$ matrix enforcing a boundary condition at $0$ \cite{BenDab-LMP-94} (note that the fact that $\b{M}$ is unitary implies that the density $\tilde{\bm{\psi}}^\dagger \tilde{\bm{\psi}}$ of any state $\tilde{\bm{\psi}} \in \text{Dom}(\b{D})$ is continuous at $x=0$). Different choices for $\b{M}$ are possible. As in Refs.~\onlinecite{SubBha-JPC-72,Lap-AJP-83,FilLorBan-JPA-12}, we choose
\begin{eqnarray}
\b{M} &=& \left(\begin{array}{cc}
\cos \theta    & \i \sin \theta\\
\i \sin \theta & \cos \theta\\
\end{array}\right),
\label{1stboundarycond}
\end{eqnarray}
with $\tan(\theta/2)=\l=Z/(2c)$. 
This boundary condition can also be obtained by integrating Eq.~(\ref{DiraceqH}) around $x=0$ and formally defining $\int_{0^-}^{0^+} \delta(x) \tilde{\bm{\psi}}(x) \d x = (1/2) \left[ \tilde{\bm{\psi}}(0^+) + \tilde{\bm{\psi}}(0^-) \right]$~\cite{SubBha-JPC-72,Lap-AJP-83}, or, more rigorously, using Colombeau's generalized theory of distributions allowing one to give a meaning to the distribution product $\delta(x) \tilde{\bm{\psi}}(x)$~\cite{FilLorBan-JPA-12}. Let us mention that another boundary condition that has also been used~\cite{CalKiaNog-AJP-87,CouNog-PRA-87,MckSte-PRC-87,GuiMunPirSan-FP-19} has the same form as Eq.~(\ref{1stboundarycond}) but with $\theta$ replaced by $\theta'=2\l$. The latter boundary condition can be obtained by considering the zero-width limit of a square-well potential~\cite{CalKiaNog-AJP-87,MckSte-PRC-87}. 

Note that the 3D hydrogen-like Dirac Hamiltonian with Coulomb potential has a unique self-adjoint extension for $Z \leq \sqrt{3}c/2$ and many self-adjoint extensions for $Z > \sqrt{3}c/2$ (see, e.g., Refs.\onlinecite{EstLewSer-RMI-19,Est-CRP-20}). The situation for the present 1D model is thus worse in the sense that the 1D hydrogen-like Dirac Hamiltonian with delta potential has many self-adjoint extensions as soon as $Z>0$. A strong motivation for using the particular self-adjoint extension determined by Eq.~(\ref{1stboundarycond}) is that it is the self-adjoint extension that seems to be numerically obtained when working in a basis of smooth functions such as Hermite functions or plane waves. This point will be further discussed in a forthcoming work.

The Hamiltonian $\b{D}$ has a single bound state with positive energy~\cite{Lap-AJP-83,CouNog-PRA-87,NogBea-EL-86}
\begin{eqnarray}
\tilde{\varepsilon}_1 = mc^2 \frac{1-\l^2}{1+\l^2},
\label{epsilon0tilde}
\end{eqnarray}
and eigenfunction
\begin{eqnarray}
\tilde{\bm{\psi}}_1(x) = A
\left(\begin{array}{c}
1\\
\i \l \sgn(x)
\end{array}\right) e^{-\kappa |x|},
\label{psi0tilde}
\end{eqnarray}
where $\sgn$ is the sign function, $\kappa = 2 mc \l/(1+\l^2)$, and $A=\sqrt{\kappa/(1+\l^2)}$. 
In the non-relativistic limit ($c\to\infty$), we have $\lambda \to 0$ and $\kappa \to mZ$, so we properly recover the bound-state eigenfunction $\sqrt{mZ} e^{-\kappa |x|}$ of the non-relativistic 1D hydrogen-like atom~\cite{Fro-JCP-56,TraGinTou-JCP-22}. In this limit, the bound-state energy has the expansion
\begin{eqnarray}
\tilde{\varepsilon}_1 = mc^2 - \frac{mZ^2}{2} + \frac{mZ^4}{8c^2} + O\left(\frac{1}{c^4} \right),
\label{epsilon0tildelargec}
\end{eqnarray} 
where $- mZ^2/2$ is the non-relativistic bound-state energy and we notice that the leading relativistic correction $mZ^4/8c^2$ has an opposite sign compared to the case of the ground-state energy of the standard 3D Dirac hydrogen-like atom with Coulomb potential (see, e.g., Ref.~\onlinecite{ReiWol-BOOK-09}).

Beside the bound state, the Hamiltonian $\b{D}$ has also a continuous energy spectrum $(-\infty,-mc^2]\cup [mc^2,+\infty)$. The generalized eigenfunctions associated with the positive energy $\varepsilon_k= \sqrt{k^2 c^2 + m^2 c^4}$ are~\cite{NogBea-EL-86}
\begin{subequations}
\begin{gather}
\tilde{\bm{\psi}}_{+,k}^\g (x)= A_k \left(\!\begin{array}{c}
\cos(k |x| + \delta_k^{+})\\
\i s_k \sgn(x) \sin(k|x|+\delta_k^{+})
\end{array}\!\right),  \; k \in (0,+\infty),\\
\tilde{\bm{\psi}}_{+,k}^\u (x)= A_k \left(\!\begin{array}{c}
\sgn(x) \sin(k|x|+\delta_k^{-})\\
-\i s_k \cos(k |x| + \delta_k^{-})
\end{array}\!\right), \; k \in (0,+\infty),
\end{gather}
\label{electronicWF}
\end{subequations}
and the generalized eigenfunctions associated with the negative energy $-\varepsilon_k$ are
\begin{subequations}
\begin{gather}
\tilde{\bm{\psi}}_{-,k}^\g (x)= A_k \left(\!\begin{array}{c}
\i s_k \cos(k |x| - \delta_k^{-})\\
\sgn(x) \sin(k|x|-\delta_k^{-})
\end{array}\!\right), \; k \in (0,+\infty),\\
\tilde{\bm{\psi}}_{-,k}^\u (x)= A_k \left(\!\begin{array}{c}
- \i s_k \sgn(x) \sin(k|x|-\delta_k^{+})\\
\cos(k |x| - \delta_k^{+})
\end{array}\!\right), \; k \in (0,+\infty),
\end{gather}
\label{positronicWF}
\end{subequations}
where $\tan \delta_k^{\pm} = \l (\varepsilon_k \pm mc^2)/(kc)$. In the non-relativistic limit, we have $\delta_k^{+} \to m Z/k$ and $\delta_k^{-} \to 0$, and we properly recover the continuum eigenstates of non-relativistic 1D hydrogen-like atom~\cite{Bro-AJP-75}.

\section{Second-quantized effective quantum electrodynamics}
\label{sec:secondquantized}

In this section, we start by considering a finite-dimensional approximation to the Hilbert space of the first-quantized one-electron theory, e.g. $\calh^{L,\Lambda} = \calh_\text{s}^{L,\Lambda} \otimes \mathbb{C}^2$ where the spatial part can be chosen as~\cite{HaiLewSerSol-PRA-07}
\begin{eqnarray}
\calh_\text{s}^{L,\Lambda} = \text{span}\left( x \in \Omega_L \mapsto e^{\i k x } \; | \; k \in \frac{2\pi \mathbb{Z}}{L}, |k|\leq \Lambda \right),\;
\end{eqnarray}
corresponding to an electron on the interval $\Omega_L = (-L/2,L/2)$ with maximal momentum $\Lambda$. The infrared (IR) cutoff $L$ is convenient to discretize the generalized continuum eigenfunctions and thus write sums over these eigenfunctions instead of integrals. The ultraviolet (UV) cutoff $\Lambda$ is necessary to avoid divergences of some quantities such as total energies. We stress that we introduce these cutoffs only for formally writing the second-quantized theory, but we do not actually solve the Dirac equation with these cutoffs. Ultimately, we will take the limits $L \to \infty$ and $\Lambda \to \infty$ of non-diverging physical quantities and we will thus use the solutions of the Dirac equation in the infinite-dimensional Hilbert space $\calh$ obtained in Section~\ref{sec:firstquantized}.

\subsection{Electron-positron Hamiltonian in Fock space}

On a finite-dimensional space, the solutions of the free-electron Dirac equation in Eq.~(\ref{Dpsi=epspsi}) form a discrete finite set of $M=M_{\PS}+M_{\NS}$ orbitals, which can be partitioned into a set of $M_{\PS}$ positive-energy states (PS) $\{ \bm{\psi}_p\}_{p\in \PS}$ and a set of $M_{\NS}$ negative-energy states (NS) $\{ \bm{\psi}_p\}_{p\in \NS}$. 

We can now introduce the relativistic fermionic Fock space ${\cal F}$ which is just a $2^{M}$-dimensional complex Hilbert space, i.e. ${\cal F}\cong \mathbb{C}^{2^M}$ where $\cong$ means ``isomorphic to''. More operationally, it is written as a direct sum
\begin{eqnarray}
{\cal F} = \bigoplus_{(n,m)=(0,0)}^{(M_{\PS},M_{\NS})} {\cal H}^{(n,m)},
\end{eqnarray}
where ${\cal H}^{(n,m)}$ represents the Hilbert space of $n$ free electrons and $m$ free positrons which is defined in a second-quantization formalism as follows. We introduce electron annihilation operators $\{ \hat{b}_p \}_{p\in \PS}$ and positron annihilation operators $\{ \hat{d}_p \}_{p\in \NS}$ acting in the Fock space, and their adjoint creation operators $\{ \hat{b}_p^\dagger \}_{p\in \PS}$ and $\{ \hat{d}_p^\dagger \}_{p\in \NS}$, respectively, such that the anticommutator of any two of these operators is zero except for 
\begin{eqnarray}
\forall p,q \in \PS, \{\hat{b}_p,\hat{b}_q^\dagger\}=\delta_{p,q}\; \text{and}\; \forall p,q \in \NS, \{\hat{d}_p,\hat{d}_q^\dagger\}=\delta_{p,q}. \;\;
\end{eqnarray}
We also introduce the free vacuum state $\ket{0} \in {\cal F}$ such that
\begin{eqnarray}
\forall p \in \PS, \hat{b}_p \ket{0} = 0 \; \text{and}\; \forall p \in \NS, \hat{d}_p \ket{0} = 0.
\end{eqnarray}
The space ${\cal H}^{(n,m)}$ is spanned by the action of $n$ electron creation operators and $m$ positron creation operators on the vacuum state $\ket{0}$, in an arbitrary order,
\begin{eqnarray}
{\cal H}^{(n,m)} =\text{span}\Bigl( \hat{b}_{p_1}^\dagger \hat{b}_{p_2}^\dagger \cdots \hat{b}_{p_n}^\dagger \hat{d}_{q_1}^\dagger \hat{d}_{q_2}^\dagger \cdots \hat{d}_{q_m}^\dagger \ket{0}, \phantom{xxxx}
\nonumber\\
 p_1<p_2<\cdots <p_n \in \PS, q_1<q_2<\cdots <q_m \in \NS \Bigl).
\end{eqnarray}
In this way, the finite-dimensional Hilbert space $\calh^{L,\Lambda}$ of the first-quantized one-electron theory is reinterpreted as composed of an electronic component ${\cal H}^{(1,0)}$ and positronic component ${\cal H}^{(0,1)}$, i.e. $\calh^{L,\Lambda} \cong {\cal H}^{(1,0)} \oplus {\cal H}^{(0,1)}$ with the mapping $\bm{\psi}_p \to b_{p}^\dagger \ket{0}$ for $p \in \PS$ and $\bm{\psi}_p \to d_{p}^\dagger \ket{0}$ for $p \in \NS$. In this sense, $\calh^{L,\Lambda}$ can be considered as a subspace of the Fock space. Note that, even though we do not include spin degrees of freedom in our model, we nevertheless use a fermionic Fock space, similarly to what is done in spinless fermion models (see, e.g., Ref.~\onlinecite{PeiWhiAff-PRB-09}).

Acting in the Fock space, we define now the Dirac field operator $\hat{\bm{\psi}}(x)$ at a fixed point $x \in \Omega_L$,
\begin{eqnarray}
\hat{\bm{\psi}}(x) = \sum_{p\in \PS} \bm{\psi}_p(x) \; \hat{b}_p + \sum_{p\in \NS} \bm{\psi}_p(x) \; \hat{d}_p^\dagger,
\end{eqnarray}
where $\{\bm{\psi}_p\}_{p\in \PS \cup \NS}$ are the eigenfunctions of the free-electron Dirac equation. The Dirac field operator is an operator-valued two-component row vector, i.e $\hat{\bm{\psi}}(x) \in {\cal L}({\cal F},{\cal F})^{2\times1}$ where ${\cal L}({\cal F},{\cal F})$ is the space of linear operators from ${\cal F}$ to ${\cal F}$. We also define the one-particle density-matrix operator $\hat{\b{n}}_1(x,x') \in {\cal L}({\cal F},{\cal F})^{2\times 2}$ at points $x$ and $x'$,
\begin{eqnarray}
\hat{\b{n}}_1(x,x') = {\cal N}[ \hat{\bm{\psi}}^\dagger(x') \otimes \hat{\bm{\psi}}(x)],
\label{OPdensitymatrix}
\end{eqnarray}
and the pair density-matrix operator $\hat{\b{n}}_2(x_1,x_2)\in {\cal L}({\cal F},{\cal F})^{4\times 4}$ at points $x_1$ and $x_2$,
\begin{eqnarray}
\hat{\b{n}}_2(x_1,x_2) = - {\cal N}[ \hat{\bm{\psi}}^\dagger(x_1) \otimes \hat{\bm{\psi}}^\dagger(x_2) \otimes \hat{\bm{\psi}}(x_1) \otimes \hat{\bm{\psi}}(x_2)],
\label{n2hatx1x2}
\end{eqnarray}
where $\otimes$ designates here the tensor product (also called Kronecker product or matrix direct product, see Appendix~\ref{app:tensor}) and ${\cal N}[ ...]$ designates normal ordering of the elementary creation and annihilation operators $\hat{b}_{p}^\dagger$, $\hat{b}_p$, $\hat{d}_{p}^\dagger$, $\hat{d}_p$ associated with the free vacuum state $\ket{0}$. We recall that normal ordering of a string of creation and annihilation operators means performing anticommutations of these elementary operators to put all the annihilation operators to the right of the creation operators. Note that, in Eq.~(\ref{n2hatx1x2}), the unusual order of the field operators is due to the use of the tensor product, but the minus sign makes the matrix elements of $\hat{\b{n}}_2(x_1,x_2)$ consistent with the definition given in Ref.~\onlinecite{Tou-SPC-21}.

The normal-ordered electron-positron Hamiltonian in Fock space~\cite{ChaIra-JPB-89,ChaIraLio-JPB-89,SauVis-INC-03,Tou-SPC-21} (see, also, Refs.~\onlinecite{LiuLin-JCP-13}) can then be written as
\begin{eqnarray}
\hat{H} &=& \int_{\Omega_L} \tr[ \b{D}(x) \hat{\b{n}}_1(x,x')]_{x'=x} \d x 
\nonumber\\
&&
+ \frac{1}{2} \int_{\Omega_L}\int_{\Omega_L} \Tr[ \b{w}(x_1,x_2) \hat{\b{n}}_2(x_1,x_2)] \d x_1 \d x_2,
\label{Hep}
\end{eqnarray}
where $\tr$ and $\Tr$ designate the trace for $2\times2$ and $4\times4$ matrices, respectively.
In Eq.~(\ref{Hep}), $\b{w}(x_1,x_2)$ is the $4\times 4$ two-particle interaction matrix chosen as~\cite{CouNog-PRA-87}
\begin{eqnarray}
\b{w}(x_1,x_2) = \delta(x_1-x_2) \left( \b{I}_2\otimes\b{I}_2 - \bm{\sigma}_1 \otimes \bm{\sigma}_1 \right),
\label{wx1x2}
\end{eqnarray}
where the first and second terms are the 1D analogs of the Coulomb and Breit interactions, respectively. Note that, in 3D, the Breit interaction is composed of the magnetic Gaunt term and the remaining retardation correction term (see, e.g., Ref.~\onlinecite{Tou-SPC-21}). In 1D, however, the Breit interaction exactly reduces to the Gaunt interaction. Hence, what we call the Breit interaction in the present work may as well be called Gaunt interaction. Due to the normal ordering, the Hamiltonian in Eq.~(\ref{Hep}) gives zero energy to the free vacuum state, i.e. $\bra{0}\hat{H}\ket{0}=0$. Note that, up to a constant, the Hamiltonian in Eq.~(\ref{Hep}) can equivalently be written with commutators and anticommutators of field operators~\cite{Tou-SPC-21}.

The electron-positron Hamiltonian $\hat{H}$ does not commute separately with the electron and positron number operators, $\hat{N}_\text{e} = \sum_{p\in\PS} \hat{b}_p^\dagger \hat{b}_p$ and $\hat{N}_\text{p} = \sum_{p\in\NS} \hat{d}_p^\dagger \hat{d}_p$, i.e. it does not conserve electron or positron numbers. However, the Hamiltonian $\hat{H}$ commutes with the opposite charge operator (or electron-excess number operator) $\hat{N} = \hat{N}_\text{e} - \hat{N}_\text{p}$, i.e. it conserves charge. It is therefore more relevant to decompose the Fock space into charge sectors
\begin{eqnarray}
{\cal F} = \bigoplus_{N=-M_{\NS}}^{M_{\PS}} {\cal F}_N,
\end{eqnarray}
where ${\cal F}_N$ is the Fock space sector of opposite charge $N$. For $N\geq 0$, we have ${\cal F}_N = {\cal H}^{(N,0)}\oplus{\cal H}^{(N+1,1)}\oplus\cdots\oplus {\cal H}^{(M_\PS,M_\NS-N)}$, and, for $N \leq 0$, we have ${\cal F}_N = {\cal H}^{(0,|N|)}\oplus{\cal H}^{(1,|N|+1)}\oplus \cdots\oplus{\cal H}^{(M_\PS-|N|,M_\NS)}$. The lowest energy for $N \geq 0$ negative charges is then obtained by the following  minimization
\begin{eqnarray}
E_N = \min_{\ket{\Psi}\in {\cal W}_N} \bra{\Psi} \hat{H} \ket{\Psi} = \bra{\Psi_N} \hat{H} \ket{\Psi_N},
\label{ENminPsi}
\end{eqnarray}
where ${\cal W}_N = \Big\{ \ket{\Psi} \in {\cal F}_N \;|\; \braket{\Psi}{\Psi}=1 \Big\}$ is the space of normalized Fock states with $N$ negative charges and $\ket{\Psi_N}$ is a minimizing state. The existence of the minimum in Eq.~(\ref{ENminPsi}) is guaranteed by the fact that we work in a finite-dimensional setting, and in particular by the UV cutoff which prevents any collapse to infinitely negative energy. Of particular interest is the correlated vacuum energy $E_0=\bra{\Psi_0} \hat{H} \ket{\Psi_0}$, which is the ground-state energy of the Hamiltonian $\hat{H}$ and is necessarily negative (since $E_0\leq \bra{0}\hat{H}\ket{0}=0$). Also of interest is the lowest energy for one negative charge $E_1=\bra{\Psi_1} \hat{H} \ket{\Psi_1}$, corresponding to a 1D hydrogen-like atom including effective QED electron-positron effects.

Like in the 3D case with Coulomb interaction~\cite{HaiLewSerSol-PRA-07}, we expect that the energies $E_N$ remain finite in the IR limit $L\to\infty$ but diverge to $-\infty$ in the UV limit $\Lambda\to\infty$. However, we speculate that the relative energies with respect to the correlated vacuum energy
\begin{eqnarray}
{\cal E}_N = E_N - E_0
\end{eqnarray}
remain finite as $L\to \infty$ and $\Lambda \to \infty$.  The first quantity of interest is thus ${\cal E}_1$, i.e. the ground-state energy of a 1D hydrogen-like atom including effective QED electron-positron effects with respect to the correlated vacuum energy. In Section~\ref{sec:firstorderh}, we will show that the first-order perturbative estimate of ${\cal E}_1$ with respect to the two-particle interaction remains indeed finite as $L\to \infty$ and $\Lambda \to \infty$. 

We note that, in the limit of a zero external nuclear potential (i.e. $Z=0$), Eq.~(\ref{Hep}) reduces to the Hamiltonian of the massive Thirring model (up to a possible different choice of normal ordering) which can be diagonalized exactly with the Bethe ansatz~\cite{BerTha-PRL-79,BerTha-PRD-79,Tha-RMP-81,FujSekYam-AP-97}. However, for $Z\neq 0$, we have to resort to approximations.

\subsection{Hartree-Fock approximation}

In the Hartree-Fock (HF) approximation, the lowest energy for $N \geq 0$ negative charges is approximated as
\begin{eqnarray}
E_N^\HF = \min_{\ket{\Phi}\in {\cal S}_N} \bra{\Phi} \hat{H} \ket{\Phi},
\label{ENHFmin}
\end{eqnarray}
where the search is restricted to the manifold of $N$-electron single-determinant states ${\cal S}_N \subset {\cal F}_N$,
\begin{eqnarray}
{\cal S}_N = \left\{ \ket{\Phi} = e^{\hat{\kappa}(\bm{\kappa})} \hat{b}_1^\dagger \hat{b}_2^\dagger \cdots \hat{b}_N^\dagger \ket{0} \; | \; \bm{\kappa} \in \mathbb{C}^{M\times M}, \bm{\kappa}^\dagger =- \bm{\kappa}\right\},
\end{eqnarray}
where $e^{\hat{\kappa}(\bm{\kappa})}$ is a unitary operator in Fock space performing an orbital rotation (corresponding to a Bogoliubov transformation mixing electron annihilation operators $\hat{b}_p$ and positron creation operators $\hat{d}_p^\dagger$)~\cite{ChaIra-JPB-89,ChaIraLio-JPB-89,SauVis-INC-03,DyaFae-BOOK-07,OhsYam-IJQC-01,Ohs-ARX-01,Tou-SPC-21} with the anti-Hermitian operator $\hat{\kappa}(\bm{\kappa})$
\begin{eqnarray}
\hat{\kappa}(\bm{\kappa}) 
&=& \sum_{p \in \PS} \sum_{q \in \PS} \kappa_{p,q} \hat{b}_p^\dagger \hat{b}_q + \sum_{p\in\PS} \sum_{q\in\NS} \kappa_{p,q} \hat{b}_p^\dagger \hat{d}_q^\dagger 
\nonumber\\
&& + \sum_{p\in\NS} \sum_{q\in\PS} \kappa_{p,q} \hat{d}_p \hat{b}_q + \sum_{p\in\NS} \sum_{q\in\NS} \kappa_{p,q} \hat{d}_p \hat{d}_q^\dagger,
\label{kappa}
\end{eqnarray}
with the orbital rotation parameters $\kappa_{p,q}$ being the elements of the anti-Hermitian matrix $\bm{\kappa}$. The operator $e^{\hat{\kappa}(\bm{\kappa})}$ generates new creation operators $\hat{\tilde{b}}_p^\dagger = e^{\hat{\kappa}(\bm{\kappa})} \hat{b}_p^\dagger e^{-\hat{\kappa}(\bm{\kappa})}$ and a new polarized (or dressed) vacuum state $\ket{\tilde{0}} = e^{\hat{\kappa}(\bm{\kappa})} \ket{0}$, such that a single-determinant state $\ket{\Phi}\in {\cal S}_N$ can be written as
\begin{eqnarray}
\ket{\Phi} &=& \hat{\tilde{b}}_1^\dagger \hat{\tilde{b}}_2^\dagger \cdots \hat{\tilde{b}}_N^\dagger \ket{\tilde{0}},
\end{eqnarray}
and the corresponding new orbitals are obtained from the original ones via the unitary matrix $\b{U}=e^{\bm{\kappa}}$
\begin{eqnarray}
\forall p \in \PS\cup\NS, \; \tilde{\bm{\phi}}_p(x) = \sum_{q\in\PS\cup\NS} \bm{\psi}_q(x) \; \b{U}_{q,p}.
\end{eqnarray}

Putting the Hamiltonian of Eq.~(\ref{Hep}) in normal ordering with respect to the single-determinant state $\Phi$ (see Refs.~\onlinecite{ChaIra-JPB-89,Tou-SPC-21}) leads to the expression of the HF energy as
\begin{eqnarray}
E_N^\HF &=& \int_{\Omega_L} \tr[ \b{D}(x) \b{n}_1^\HF(x,x')]_{x'=x} \d x 
\nonumber\\
&&
+ \frac{1}{2} \int_{\Omega_L}\int_{\Omega_L} \Tr[ \b{w}(x_1,x_2) \b{n}_2^\HF(x_1,x_2)] \d x_1 \d x_2,
\label{}
\end{eqnarray}
where $\b{n}_1^\HF(x,x') = \bra{\Phi} \hat{\b{n}}_1(x,x') \ket{\Phi}$ is the HF one-particle density matrix which can be written as
\begin{eqnarray}
\b{n}_1^\HF(x,x') = \b{n}_1^{\HF,\el}(x,x') + \b{n}_1^{\HF,\vp}(x,x'),
\label{n1HF}
\end{eqnarray}
including the contribution from the occupied electronic (el) orbitals
\begin{eqnarray}
\b{n}_1^{\HF,\el}(x,x') = \sum_{i=1}^N \tilde{\bm{\phi}}_i(x) \tilde{\bm{\phi}}_i^\dagger(x'),
\label{}
\end{eqnarray}
and the vacuum-polarization (vp) contribution
\begin{eqnarray}
\b{n}_1^{\HF,\vp}(x,x') = \sum_{p \in \NS} \tilde{\bm{\phi}}_p(x) \tilde{\bm{\phi}}_p^\dagger(x') - \sum_{p \in \NS} \bm{\psi}_p(x) \bm{\psi}_p^{\dagger}(x'),
\label{}
\end{eqnarray}
and $\b{n}_2^\HF(x_1,x_2) = \bra{\Phi} \hat{\b{n}}_2(x_1,x_2) \ket{\Phi}$ is the HF pair-density matrix
\begin{eqnarray}
\b{n}_2^\HF(x_1,x_2) &=& \b{n}_1^\HF(x_1,x_1) \otimes \b{n}_1^\HF(x_2,x_2) 
\nonumber\\
&&-\b{X} \; \left( \b{n}_1^\HF(x_2,x_1) \otimes \b{n}_1^\HF(x_1,x_2) \right),
\label{}
\end{eqnarray}
where $\b{X}$ is the permutation matrix 
\begin{eqnarray}
\b{X} = \left(\begin{array}{cccc}
1 & 0 & 0 & 0\\
0 & 0 & 1 & 0\\
0 & 1 & 0 & 0\\
0 & 0 & 0 & 1\\
\end{array}\right),
\end{eqnarray}
which exchanges the second and third lines in the matrix it multiplies on the right.

The stationary condition corresponding to the minimization in Eq.~(\ref{ENHFmin}) leads the following HF equations which determine the HF orbitals $\{ \tilde{\bm{\phi}}_p \}_{p\in \PS \cup \NS}$ and HF orbital energies $\{\tilde{\epsilon}_p\}_{p\in \PS \cup \NS}$
\begin{eqnarray}
\left( \b{D}(x) + \b{v}_\H(x)  \right) \tilde{\bm{\phi}}_p(x) 
+ \int_{\Omega_L} \!\! \b{v}_\x(x,x') \tilde{\bm{\phi}}_p(x') \d x'
= \tilde{\epsilon}_p \tilde{\bm{\phi}}_p(x),\;\;
\label{}
\end{eqnarray}
with the local $2\times2$ Hartree potential
\begin{eqnarray}
\b{v}_\H(x_1) = \int_{\Omega_L} \Tr_2[ \b{w}(x_1,x_2) \; (\b{I}_2 \otimes \b{n}_1^\HF(x_2,x_2)) ] \d x_2,
\label{vH}
\end{eqnarray}
and the non-local $2\times2$ exchange potential
\begin{eqnarray}
\b{v}_\x(x_1,x_2) = -\Tr_2[ \b{w}(x_1,x_2) \; \b{X} (\b{I}_2 \otimes \b{n}_1^\HF(x_1,x_2)) ],
\label{vX}
\end{eqnarray}
where $\Tr_2$ designates the partial trace with respect to the second particle (see Appendix~\ref{app:tensor}). 

As for the exact energies, we expect the HF total energies $E_N^\HF$ to diverge to $-\infty$ in the UV limit $\Lambda\to\infty$. It is then natural to consider the HF relative energies with respect to the HF vacuum energy
\begin{eqnarray}
{\cal E}_N^\HF = E_N^\HF - E_0^\HF,
\end{eqnarray}
which should remain finite as $L\to \infty$ and $\Lambda \to \infty$. Unfortunately, even for the present relatively simple 1D model, the HF equations cannot be solved exactly, even for $N=0$. If the vacuum-polarization density matrix $\b{n}_1^{\HF,\vp}(x,x')$ is neglected in Eq.~(\ref{n1HF}), the present HF equations reduce to standard non-QED relativistic HF equations. In particular, for $N=1$, the latter equations simply reduce to the hydrogen-like Dirac equation [Eq.~(\ref{DiraceqH})].

\subsection{First-order perturbation theory}

Since the HF equations cannot be solved exactly, we consider instead a perturbation theory with respect to the two-particle interaction $\b{w}(x_1,x_2)$ [Eq.~(\ref{wx1x2})]. 

Instead of the HF one-particle density matrix in Eq.~(\ref{n1HF}), we thus consider the zeroth-order one-particle density matrix
\begin{eqnarray}
\b{n}_1^{}(x,x') = \b{n}_1^\el(x,x') + \b{n}_1^\vp(x,x'),
\label{n10}
\end{eqnarray}
with the contribution from the occupied electronic orbitals
\begin{eqnarray}
\b{n}_1^\el(x,x') = \sum_{i=1}^N \tilde{\bm{\psi}}_i(x) \tilde{\bm{\psi}}_i^\dagger(x'),
\label{}
\end{eqnarray}
and the vacuum-polarization contribution
\begin{eqnarray}
\b{n}_1^\vp(x,x') = \sum_{p \in \NS} \tilde{\bm{\psi}}_p(x) \tilde{\bm{\psi}}_p^\dagger(x') - \sum_{p \in \NS} \bm{\psi}_p(x) \bm{\psi}_p^{\dagger}(x'),
\label{n1vpxxp}
\end{eqnarray}
where the zeroth-order orbitals $\{ \tilde{\bm{\psi}}_p \}_{p\in \PS \cup \NS}$ are the 1D hydrogen-like orbitals (determined in Section~\ref{sec:hydrogenlike} in the limits $L\to\infty$ and $\Lambda\to\infty$). The zeroth-order energy is then
\begin{eqnarray}
E_N^{(0)} &=& \int_{\Omega_L} \tr[ \b{D}(x) \b{n}_1^{}(x,x')]_{x'=x} \d x,
\label{}
\end{eqnarray}
and the zeroth-order relative energy with respect to the vacuum is
\begin{eqnarray}
{\cal E}_N^{(0)} &=& E_N^{(0)} - E_0^{(0)}
\nonumber\\
                 &=& \int_{\Omega_L} \tr[ \b{D}(x) \b{n}_1^\el(x,x')]_{x'=x} \d x.
\label{EN0}
\end{eqnarray}

Let us move now to the first-order energy correction which is
\begin{eqnarray}
E_N^{(1)} &=& 
\frac{1}{2} \int_{\Omega_L}\int_{\Omega_L} \Tr[ \b{w}(x_1,x_2) \b{n}_2^{}(x_1,x_2)] \d x_1 \d x_2,
\label{}
\end{eqnarray}
where $\b{n}_2^{}(x_1,x_2)$ is the zeroth-order pair-density matrix
\begin{eqnarray}
\b{n}_2^{}(x_1,x_2) &=& \b{n}_1^{}(x_1,x_1) \otimes \b{n}_1^{}(x_2,x_2) 
\nonumber\\
&&-\b{X} \; \left( \b{n}_1^{}(x_2,x_1) \otimes \b{n}_1^{}(x_1,x_2) \right).
\label{}
\end{eqnarray}
The first-order relative correction is thus
\begin{eqnarray}
{\cal E}_N^{(1)} &=& E_N^{(1)} - E_0^{(1)}
\nonumber\\
                 &=& \frac{1}{2} \int_{\Omega_L}\int_{\Omega_L} \Tr[ \b{w}(x_1,x_2) \Delta \b{n}_2^{}(x_1,x_2)] \d x_1 \d x_2,
\label{}
\end{eqnarray}
where
\begin{eqnarray}
\Delta \b{n}_2^{}(x_1,x_2) &=& \b{n}_2^{}(x_1,x_2) - \b{n}_2^{\vp}(x_1,x_2),
\label{}
\end{eqnarray}
and $\b{n}_2^{\vp}(x_1,x_2)$ is the zeroth-order vacuum-polarization pair-density matrix
\begin{eqnarray}
\b{n}_2^{\vp}(x_1,x_2) &=& \b{n}_1^{\vp}(x_1,x_1) \otimes \b{n}_1^{\vp}(x_2,x_2) 
\nonumber\\
&&-\b{X} \; \left( \b{n}_1^{\vp}(x_2,x_1) \otimes \b{n}_1^{\vp}(x_1,x_2) \right).
\label{}
\end{eqnarray}
It can be decomposed as 
\begin{eqnarray}
{\cal E}_N^{(1)} &=& {\cal E}_N^{\el,(1)} + {\cal E}_N^{\vp,(1)},
\label{}
\end{eqnarray}
where ${\cal E}_N^{\el,(1)}$ is the contribution coming only from the occupied electronic orbitals 
\begin{eqnarray}
{\cal E}_N^{\el,(1)} = \frac{1}{2} \int_{\Omega_L}\int_{\Omega_L} \Tr[ \b{w}(x_1,x_2) \b{n}_2^{\el}(x_1,x_2)] \d x_1 \d x_2,
\label{ENel1}
\end{eqnarray}
with 
\begin{eqnarray}
\b{n}_2^{\el}(x_1,x_2) &=& \b{n}_1^{\el}(x_1,x_1) \otimes \b{n}_1^{\el}(x_2,x_2) 
\nonumber\\
&&-\b{X} \; \left( \b{n}_1^{\el}(x_2,x_1) \otimes \b{n}_1^{\el}(x_1,x_2) \right),
\label{}
\end{eqnarray}
and ${\cal E}_N^{\vp,(1)}$ is the contribution involving vacuum-polarization terms
\begin{eqnarray}
{\cal E}_N^{\vp,(1)} = \int_{\Omega_L}\int_{\Omega_L} \Tr[ \b{w}(x_1,x_2) \b{n}_2^{\el/\vp}(x_1,x_2)] \d x_1 \d x_2,
\label{}
\end{eqnarray}
with
\begin{eqnarray}
\b{n}_2^{\el/\vp}(x_1,x_2) &=& \b{n}_1^{\el}(x_1,x_1) \otimes \b{n}_1^{\vp}(x_2,x_2) 
\nonumber\\
&&-\b{X} \; \left( \b{n}_1^{\el}(x_2,x_1) \otimes \b{n}_1^{\vp}(x_1,x_2) \right).
\label{}
\end{eqnarray}

For the form of the two-particle interaction in Eq.~(\ref{wx1x2}), the vacuum-polarization first-order relative energy correction has four contributions
\begin{eqnarray}
{\cal E}_N^{\vp,(1)} = {\cal E}_N^{\vp,(1),\text{DC}} +{\cal E}_N^{\vp,(1),\text{XC}} + {\cal E}_N^{\vp,(1),\text{DB}} +{\cal E}_N^{\vp,(1),\text{XB}}.
\label{ENvp1}
\end{eqnarray}
The direct Coulomb-type (DC) and exchange Coulomb-type (XC) contributions are
\begin{eqnarray}
{\cal E}_N^{\vp,(1),\text{DC}} = \int_{\Omega_L} n^\el(x) n^\vp(x) \d x,
\label{ENvp1DC}
\end{eqnarray}
and
\begin{eqnarray}
{\cal E}_N^{\vp,(1),\text{XC}} = -\int_{\Omega_L} \tr[ \b{n}_1^{\el}(x) \b{n}_1^{\vp}(x) ]  \d x,
\label{ENvp1XC}
\end{eqnarray}
where we have introduced the local electronic and vacuum-polarization density matrices $\b{n}_1^{\el}(x) =  \b{n}_1^{\el}(x,x)$ and $\b{n}_1^{\vp}(x)=\b{n}_1^{\vp}(x,x)$, and the associated electronic and vacuum-polarization densities $n^\el(x) = \tr[ \b{n}_1^{\el}(x)]$ and $n^\vp(x) = \tr[ \b{n}_1^{\vp}(x)]$. Similarly, the direct Breit-type (DB) and exchange Breit-type (XB) contributions are
\begin{eqnarray}
{\cal E}_N^{\vp,(1),\text{DB}} = -\frac{1}{c^2} \int_{\Omega_L} j^\el(x) j^\vp(x) \d x,
\label{ENvp1DB}
\end{eqnarray}
and
\begin{eqnarray}
{\cal E}_N^{\vp,(1),\text{XB}} = \frac{1}{c^2} \int_{\Omega_L} \tr[ \b{j}_1^{\el}(x) \b{j}_1^{\vp}(x) ]  \d x,
\label{ENvp1XB}
\end{eqnarray}
where we have introduced the local electronic and vacuum-polarization current-density matrices $\b{j}_1^{\el}(x) =  c \bm{\sigma}_1 \b{n}_1^{\el}(x)$ and $\b{j}_1^{\vp}(x) =  c \bm{\sigma}_1 \b{n}_1^{\vp}(x)$, and the associated electronic and vacuum-polarization current densities $j^\el(x) = \tr[ \b{j}_1^{\el}(x)]$ and $j^\vp(x) = \tr[ \b{j}_1^{\vp}(x)]$.

Note that, in standard QED, the direct contributions in Eqs.~(\ref{ENvp1DC}) and~(\ref{ENvp1DB}) and the exchange contributions in Eq.~(\ref{ENvp1XC}) and~(\ref{ENvp1XB}) are often called ``vacuum polarization'' and ``self-energy'' contributions to the Lamb shift, respectively (see, e.g., Refs.~\onlinecite{MohPluSof-PR-98,SunSalSau-JCP-22}). Here, as in Ref.~\onlinecite{Tou-SPC-21}, we adopt the terminology of Ref.~\onlinecite{ChaIra-JPB-89} and use ``vacuum polarization'' to qualify both the direct and exchange contributions.

\subsection{Vacuum-polarization density}
\label{VPD}

\begin{figure*}
	\centering 
	\includegraphics[width=0.30\textwidth,angle=-90]{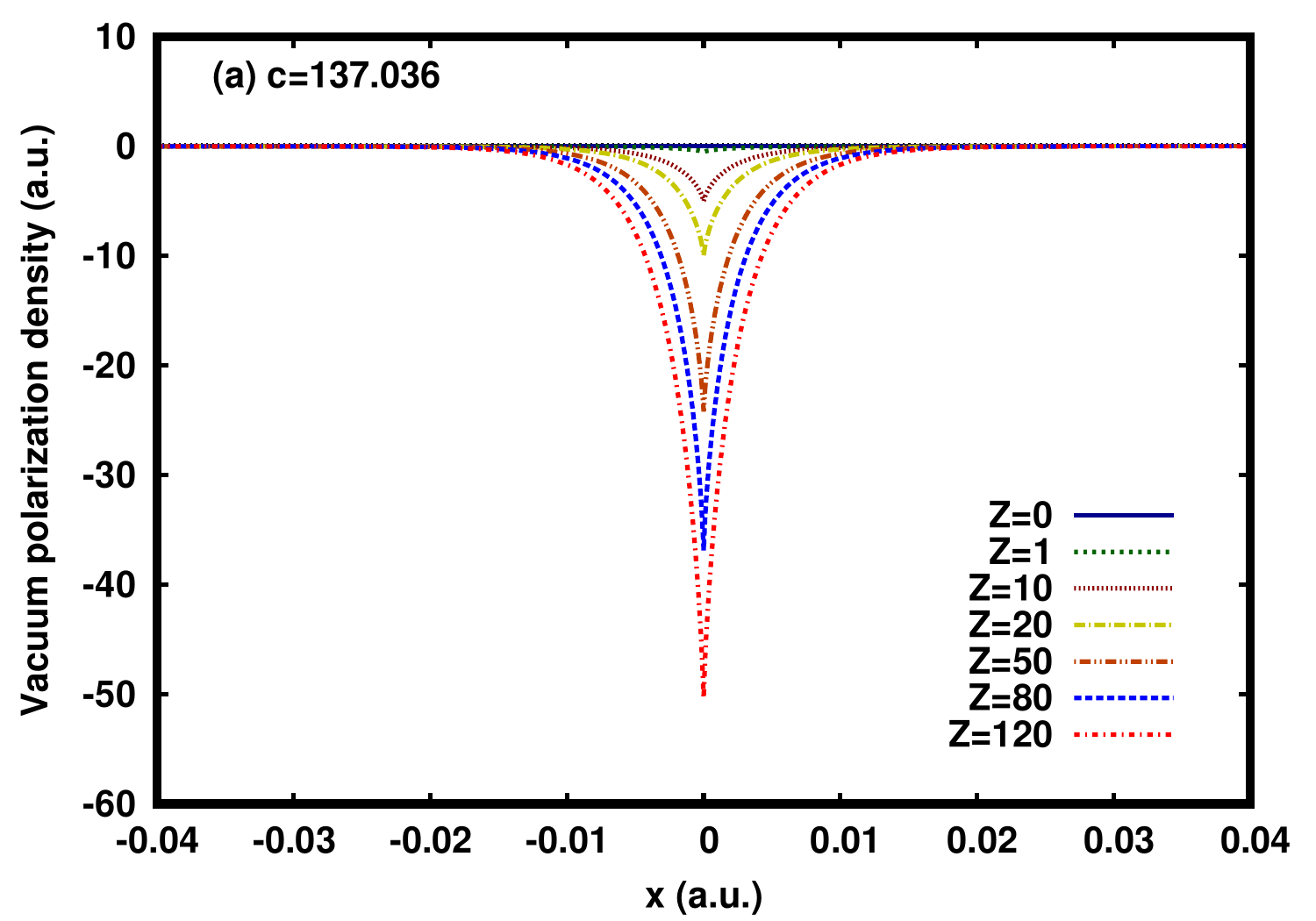}
	\includegraphics[width=0.30\textwidth,angle=-90]{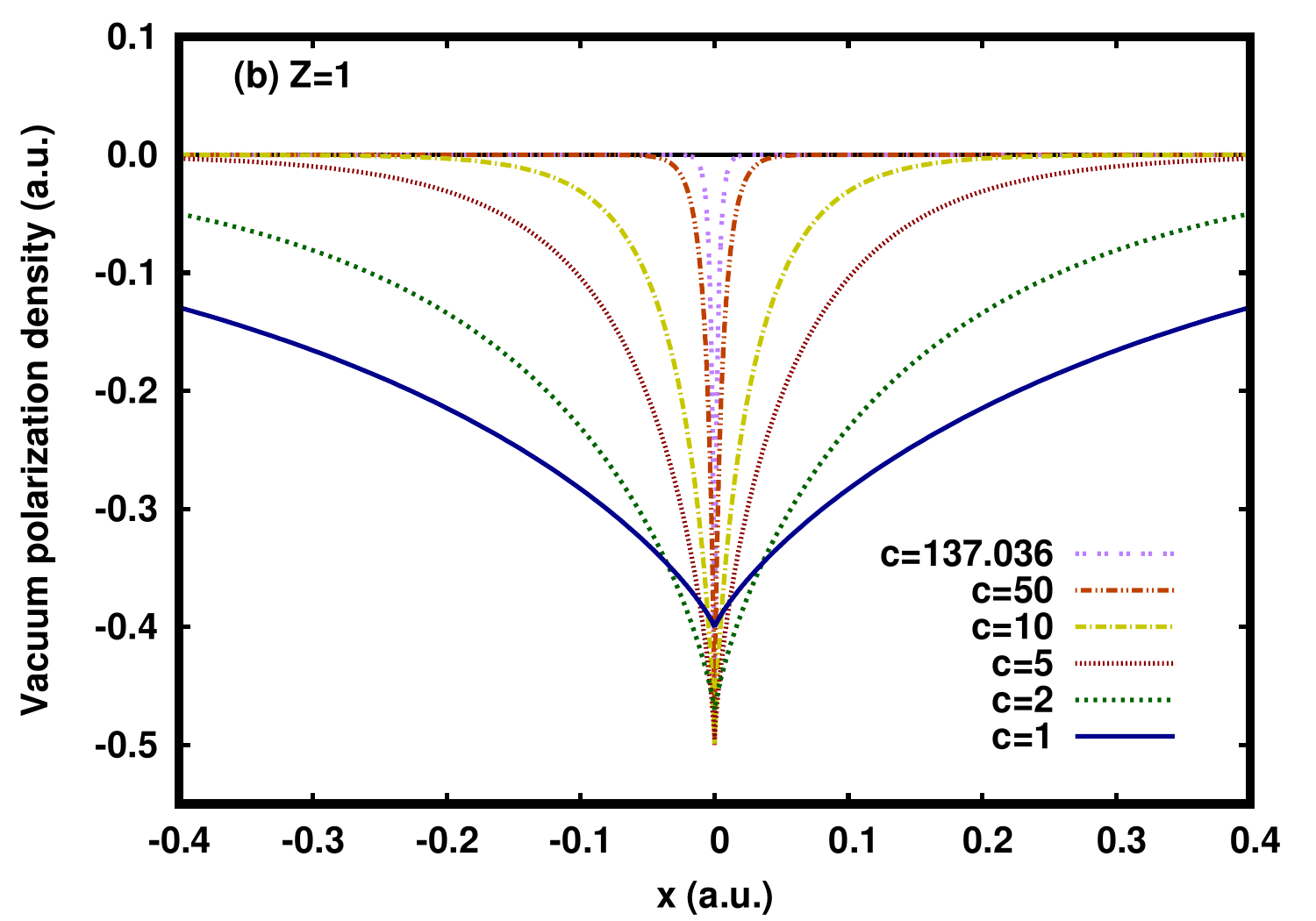}
	\caption{The vacuum-polarization density $n^{\vp}(x)$ [Eq.~(\ref{nvpx})] for (a) a fixed value of the speed of light $c = 137.036$ and different values of the nuclear charge $Z$ and (b) for a fixed value $Z=1$ and different values of $c$.}
	\label{fig:VPD}
\end{figure*}

\begin{figure}
	\centering 
	\includegraphics[width=0.30\textwidth,angle=-90]{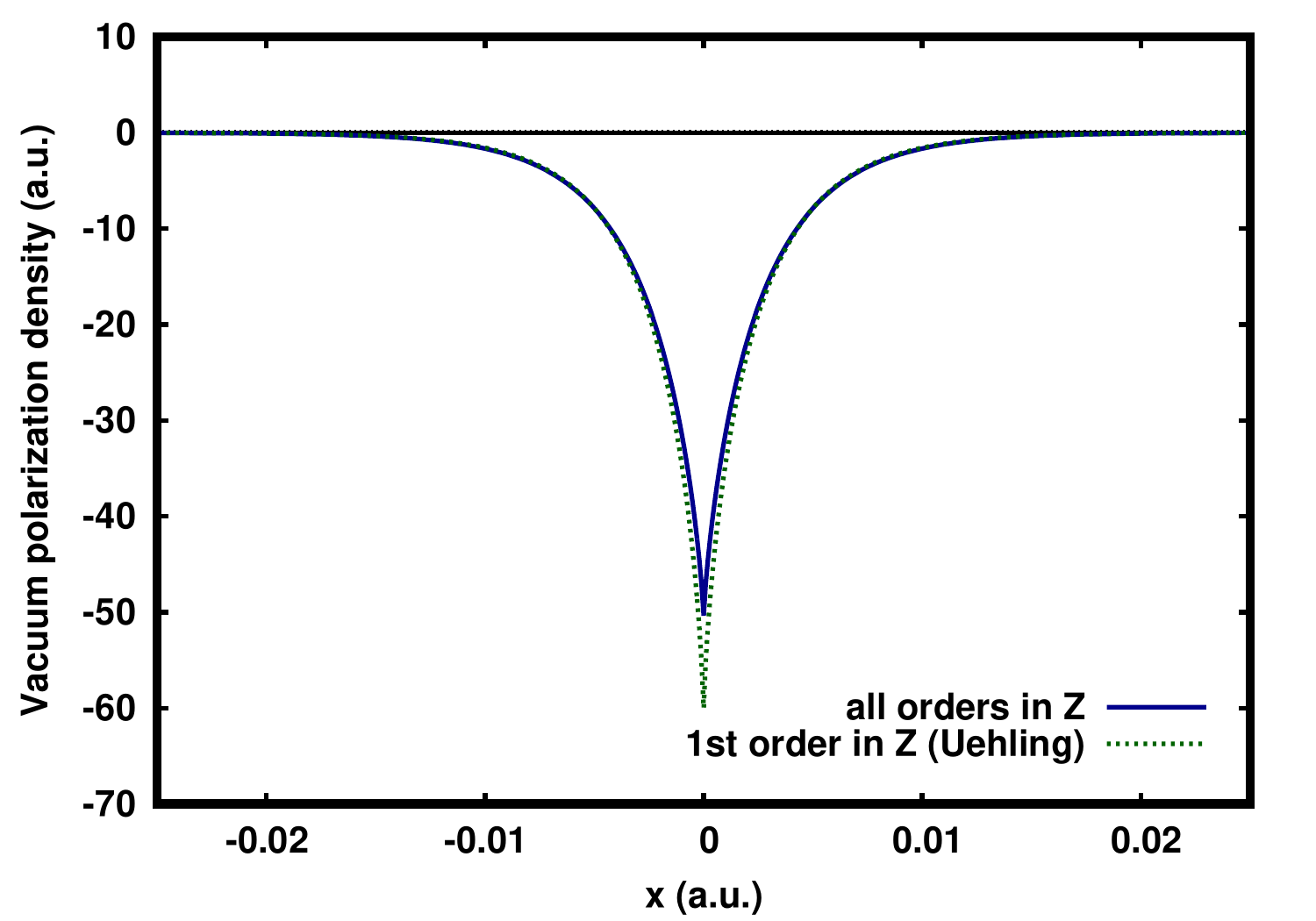}
	\caption{The vacuum-polarization density $n^{\vp}(x)$ [Eq.~(\ref{nvpx})] (all orders in $Z$) and the Uehling-type vacuum-polarization density $n^{\text{vp},(1)}(x)$ [Eq.~(\ref{nvp1xtext})] (first order in $Z$) for the speed of light $c=137.036$ and nuclear charge $Z=120$.}
	\label{fig:VPDUehling}
\end{figure}

We now calculate the vacuum-polarization density $n^{\vp}(x)$ appearing in Eq.~(\ref{ENvp1DC}) in the limit $L\to\infty$ and $\Lambda \to\infty$. We stress that this quantity is the opposite-charge vacuum-polarization density. The charge vacuum-polarization density, e.g. discussed in Ref.~\onlinecite{GreRei-BOOK-09}, is $\rho^{\vp}(x) = - n^{\vp}(x)$.

Using the negative-energy generalized eigenfunctions in Eq.~\eqref{positronicFreeWF} and \eqref{positronicWF}, we find the expressions of the local vacuum-polarization density matrix
\begin{eqnarray}
\b{n}_1^{\vp}(x)\!= - \!\int_0^{\infty} \!\! \frac{\d k}{\pi} \frac{\kappa}{k^2+\kappa^2} \frac{\varepsilon_k+mc^2}{2mc^2}
\begin{pmatrix} -s_k^2 f_k(x) & - \i s_k g_k(x) \\
 \i s_k g_k(x) & f_k(x) \end{pmatrix}, \;\;
\label{OPdensMatrixVP}	
\end{eqnarray}
where $f_k(x) = \kappa \cos(2kx) - (\tilde{\varepsilon}_1/\varepsilon_k) k \sin(2k|x|)$ and $g_k(x) = (\tilde{\varepsilon}_1/\varepsilon_k) k \cos(2kx) + \kappa \sin(2k|x|)$. We remind that $\varepsilon_k$ and $s_k$ were defined before Eq.~(\ref{electronicFreeWF}) and after Eq.~(\ref{positronicFreeWF}), respectively, and $\tilde{\varepsilon}_1$ and $\kappa$ were defined in Eq.~(\ref{epsilon0tilde}) and after Eq.~(\ref{psi0tilde}), respectively. The vacuum-polarization density is then
\begin{eqnarray}
n^{\vp}(x)= - \!\int_0^{\infty} \! \frac{\d k}{\pi} \frac{\kappa}{k^2+\kappa^2} f_k(x).
\label{nvpx}	
\end{eqnarray}
It is also easy to check that the vacuum-polarization current density vanishes, i.e $j^\vp(x)=0$.

Note that to obtain Eq.~(\ref{OPdensMatrixVP}) or Eq.~(\ref{nvpx}), we have formally taken the limits $L\to\infty$ and $\Lambda\to\infty$ in each of the two sums in Eq.~(\ref{n1vpxxp}), which gives two divergent integrals over $k$, but taking the difference of the two integrands finally gives a convergent integral over $k$. The same approach used for the alternative commutator definition of the vacuum-polarization density leads to the same result, as shown in Appendix~\ref{app:VPcommutator}. Also, the vacuum-polarization density independently calculated by Nogami and Beachey~\cite{NogBea-EL-86} on the same non-interacting model agrees numerically with the values obtained with Eq.~(\ref{nvpx}). Finally, we give in Appendix~\ref{app:green} an alternative expression of the vacuum-polarization density using a more rigorous approach based on the Green function which numerically agrees perfectly with the expression in Eq.~(\ref{nvpx}). 

The vacuum-polarization density originates from the presence of free electron-positron pairs in the polarized vacuum state due to the external potential. The vacuum-polarization density is plotted in Fig.~\ref{fig:VPD} for the physical value of the speed of light $c=137.036$ and different values of the nuclear charge $Z$, and for a fixed value of the nuclear charge $Z=1$ and different values of the speed of light $c$. For $Z=0$, the vacuum-polarization density is of course zero. For $Z \neq 0$, the vacuum-polarization density is localized around the nucleus and is always negative. At least close of the nucleus, this negative sign can be understood from Eq.~(\ref{n1vpxxp}) and the fact the external potential $-Z\delta(x)$ tends to give negative-energy eigenfunctions $\{\tilde{\bm{\psi}}_p\}_{\p\in \NS}$ with smaller probability density near the nucleus in comparison with the free negative-energy eigenfunctions $\{\bm{\psi}_p\}_{\p\in\NS}$ (for similar discussions in the standard QED case, see Ref.~\onlinecite{GreRei-BOOK-09}). As expected, the amplitude of the vacuum-polarization density increases with $Z$. As $c$ decreases, the relativistic effects increase, and the vacuum-polarization density becomes more and more extended around the nucleus.

In Appendix~\ref{app:green}, we also derive the first-order vacuum-polarization density with respect to $Z$, i.e. with respect to the external potential, as
\begin{eqnarray}
n^{\text{vp},(1)}(x) &=&  -\frac{Z m}{\pi} \int_{1}^{\infty} \d t \; \frac{e^{-2 mc |x| t}}{t\sqrt{t^2-1}},
\label{nvp1xtext}
\end{eqnarray}
which is the equivalent for the present 1D model of the Uehling vacuum-polarization density (or potential, since for a delta-interaction density and potential are identical) for the 3D hydrogen-like atom~\cite{Ueh-PR-35} (see also, e.g., Refs.~\onlinecite{GreRei-BOOK-09,IndMohSap-PRA-14}). In Eq.~(\ref{nvp1xtext}), it is manifest that the spatial range of the vacuum-polarization density is of the order of the reduced Compton wavelength $\lambdabar = 1/(mc)$. The Uehling-type vacuum-polarization density $n^{\text{vp},(1)}(x)$ is plotted in Fig.~\ref{fig:VPDUehling} for $c=137.036$ and $Z=120$. It appears to be a good approximation to the vacuum-polarization density $n^{\vp}(x)$.

In the 3D case, both for effective QED and standard QED, the calculation of the vacuum-polarization density suffers from UV divergences that require regularization, for example with a finite UV cutoff $\Lambda$, and charge renormalization to absorb the dependence on the UV cutoff (see, e.g., Refs.~\onlinecite{HaiLewSerSol-PRA-07,IndMohSap-PRA-14}). It is noteworthy that, in the present 1D model, we can obtain a finite vacuum-polarization density in the limit $\Lambda\to\infty$ without regularization and $\Lambda$-dependent charge renormalization. 

As apparent in Fig.~\ref{fig:VPD}, the integral over space of the vacuum-polarization density $n^{\vp}(x)$ is not zero. Nogami and Beachey~\cite{NogBea-EL-86} found an analytical expression for this integral
\begin{eqnarray}
\int_{-\infty}^{+\infty} \!\! n^{\text{vp}}(x) \d x &=& -\frac{2}{\pi} \arctan \left( \frac{Z}{2c} \right),
\label{intnvpx}
\end{eqnarray}
which we numerically confirmed. This means that, sufficiently far from the nucleus ($x \gg \lambdabar$), one observes a nucleus charge 
\begin{eqnarray}
Z_\text{obs} = Z + \frac{2}{\pi} \arctan \left( \frac{Z}{2c} \right).
\label{Zobs}
\end{eqnarray}
Surprisingly, the observed nucleus charge $Z_\text{obs}$ is larger than the bare nuclear charge $Z$. Thus, in contrast with 3D effective or standard QED where the bare charge is screened by the vacuum-polarization density (see, e.g., Ref.~\cite{HaiLewSerSol-PRA-07}), in the present 1D model the bare charge is (slightly) antiscreened.

However, one should not conclude from Eq.~(\ref{intnvpx}) that the vacuum state contains a fractional charge. As explained in Refs.~\onlinecite{HaiLewSer-CMP-05,HaiLewSerSol-PRA-07,GraLewSer-CMP-09}, the opposite charge of the vacuum state should be calculated as
\begin{eqnarray}
N^\text{vac} = N^\text{vac}_\text{e} - N^\text{vac}_\text{p},
\label{Nvac}
\end{eqnarray}
where $N^\text{vac}_\text{e}$ and $N^\text{vac}_\text{p}$ are the number of free electrons and free positrons, respectively, defined as
\begin{eqnarray}
N^\text{vac}_\text{e} = \int_{-\infty}^{+\infty} \int_{-\infty}^{+\infty} \tr[ \b{P}_{+}^0(x',x) \b{n}_1^{\text{vp}}(x,x') ]\d x \d x',
\label{Nvace}
\end{eqnarray}
and
\begin{eqnarray}
N^\text{vac}_\text{p} = -\int_{-\infty}^{+\infty} \int_{-\infty}^{+\infty} \tr[ \b{P}_{-}^0(x',x) \b{n}_1^{\text{vp}}(x,x') ]\d x \d x',
\label{Nvacp}
\end{eqnarray}
where $\b{P}_{+}^0(x',x)$ and $\b{P}_{-}^0(x',x)$ are the projectors on the positive-energy and negative-energy eigenfunctions of the free-particle Dirac Hamiltonian, respectively. In Appendix~\ref{app:Nvac}, we calculate $N^\text{vac}_\text{e}$ and $N^\text{vac}_\text{p}$ and find numerically $N^\text{vac}_\text{e} = N^\text{vac}_\text{p}$ (to a good precision), i.e. the vacuum state has zero (fermionic) charge
\begin{eqnarray}
N^\text{vac} = 0,
\label{}
\end{eqnarray}
as expected. If, instead of calculating $N^\text{vac}_\text{e}$ and $N^\text{vac}_\text{p}$ separately, one naively adds the integrands in Eqs.~(\ref{Nvace}) and~(\ref{Nvacp}), the projector on the full one-particle Hilbert space $\b{P}_{0}(x',x)=\b{P}_{+}^0(x',x)+\b{P}_{-}^0(x',x) = \delta(x-x') \b{I}_2$ will appear, and one will obtain the non-vanishing integral of the vacuum-polarization density
\begin{eqnarray}
\int_{-\infty}^{+\infty} \int_{-\infty}^{+\infty} \tr[ \b{P}_{0}(x',x) \b{n}_1^{\text{vp}}(x,x') ]\d x \d x' \phantom{xxxxxx}
\nonumber\\
 = \int_{-\infty}^{+\infty} \tr[ \b{n}_1^{\text{vp}}(x,x) ]\d x 
= \int_{-\infty}^{+\infty} \!\! n^{\text{vp}}(x) \d x  \neq 0.
\label{}
\end{eqnarray}
This apparent paradox comes from the fact that the vacuum-polarization density matrix $\b{n}_1^{\text{vp}}(x,x')$ is a kernel of an operator which is not trace-class, which means that the integral trace over $x$ is ill-defined. It can lead to different values depending on the way it is calculated. A very similar situation appears when calculating the screening of the charge of a defect (or impurity) by the polarization of the Fermi sea in a crystal~\cite{CanLew-ARMA-09,CanLewSto-INC-11,CanLeb-MMMAS-13}. As understood by Nogami~\cite{Nog-ARX-08}, the problem is related to the IR limit $L\to\infty$: for finite $L$ the vacuum-polarization density integrates to zero, but in the limit $L\to\infty$ there is a contribution to the vacuum-polarization density that goes uniformly to zero, so that after taking the limit $L\to\infty$ one cannot recover the total charge of the vacuum from the vacuum-polarization density. The correct zero vacuum charge is obtained by first calculating the vacuum charge for finite $L$ (which is zero) and then take the limit $L\to\infty$. Alternatively, after the limit $L\to\infty$ has been taken, the information about the zero vacuum charge can be retrieved from the vacuum-polarization density matrix $\b{n}_1^{\text{vp}}(x,x')$ via Eqs.~(\ref{Nvac})-(\ref{Nvacp}).

As noted in Ref.~\onlinecite{HaiLewSerSol-PRA-07}, in a finite-dimensional approximation, the vacuum-polarization density matrix $\b{n}_1^{\text{vp}}(x,x')$ would be trace-class, and hence the integral of the vacuum-polarization density in Eq.~(\ref{intnvpx}) would necessarily be zero. It is therefore not clear how one could estimate the observed nuclear charge in Eq.~(\ref{Zobs}) from a finite-dimensional calculation.

\subsection{First-order energy corrections for the hydrogen-like atom}
\label{sec:firstorderh}

\begin{figure*}
	\centering 
	\includegraphics[width=0.30\textwidth,angle=-90]{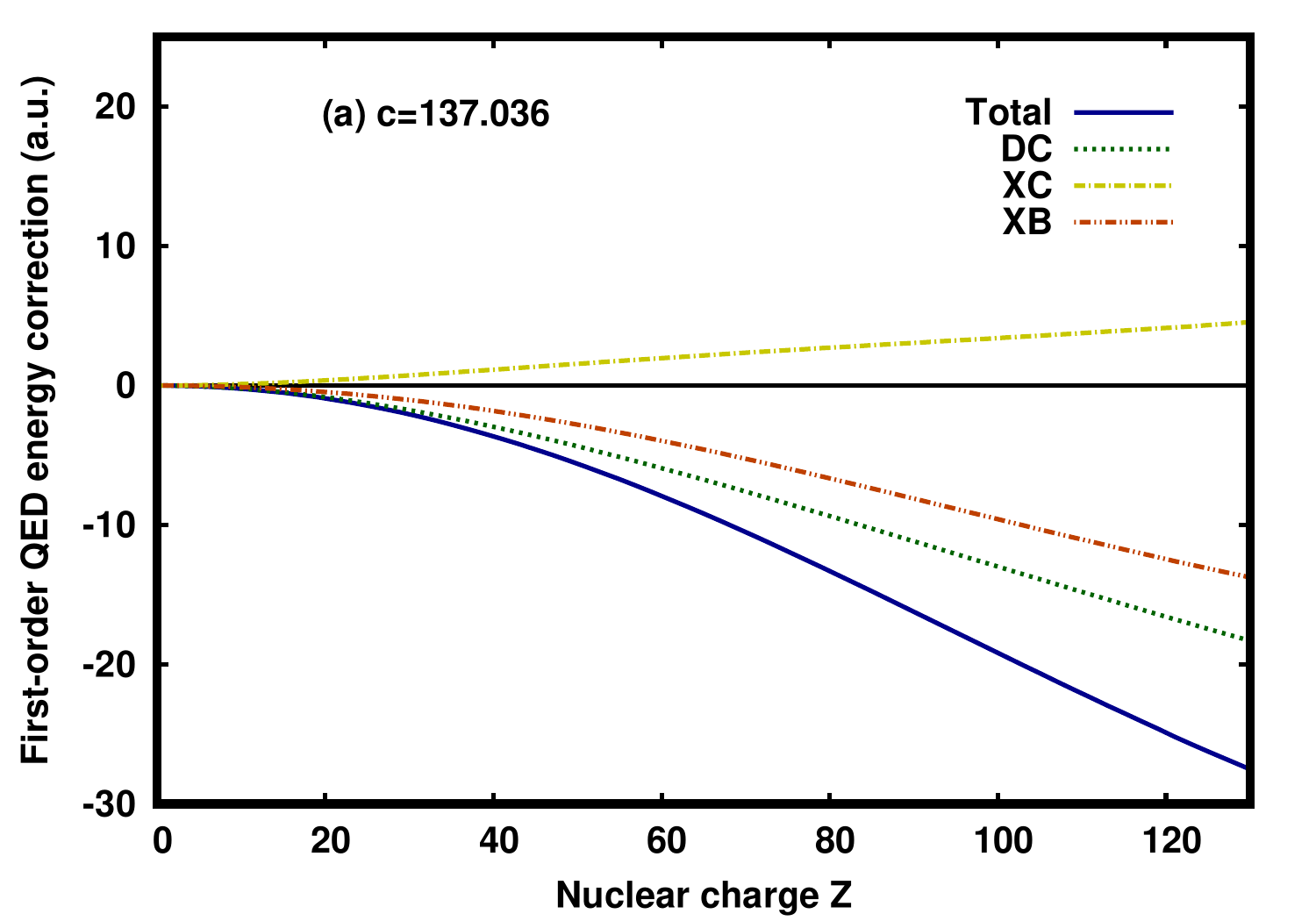}
	\includegraphics[width=0.30\textwidth,angle=-90]{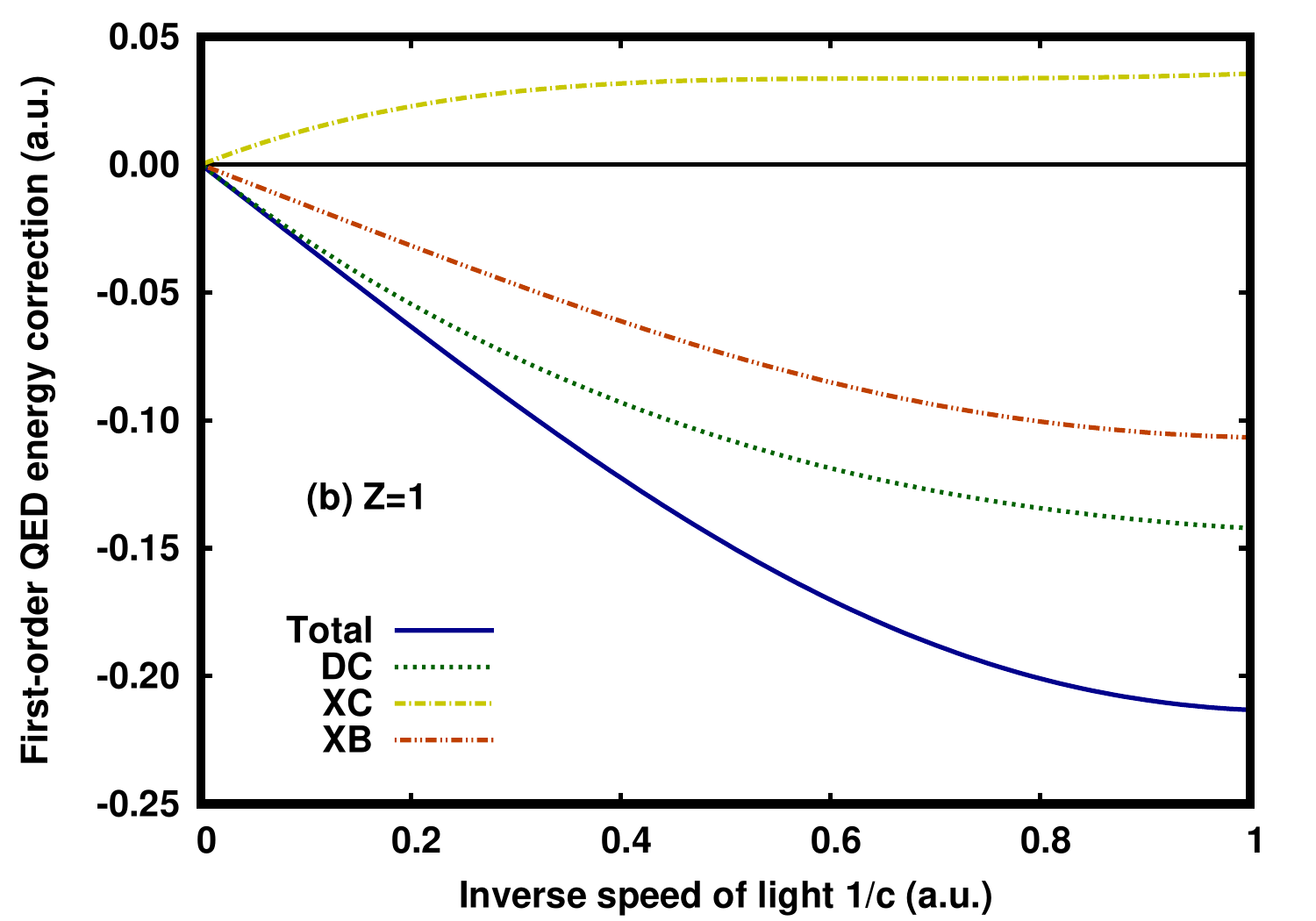}
	\caption{First-order QED vacuum-polarization correction to the bound-state energy of the 1D hydrogen-like atom (a) for $c=137.036$ and as a function of $Z$, and (b) for $Z=1$ and as a function of 1/$c$. The direct Coulomb-type (DC) [Eq.~(\ref{E1DC})], exchange Coulomb-type (XC) [Eq.~(\ref{E1XC})], and exchange Breit-type (XB) [Eq.~(\ref{E1XB})] contributions are shown, as well as the total correction.}
	\label{fig:FirstOrderQEDCorrection}
\end{figure*}

We now evaluate the first-order energy corrections in the case of the hydrogen-like atom, i.e. $N=1$, again in the limits $L\to\infty$ and $\Lambda\to\infty$.

The zeroth-order relative energy in Eq.~(\ref{EN0}) is just the bound-state orbital energy
\begin{eqnarray}
{\cal E}_1^{(0)} &=& \tilde{\varepsilon}_1,
\label{}
\end{eqnarray}
given in Eq.~(\ref{epsilon0tilde}).

The direct and exchange electronic contributions in Eq.~(\ref{ENel1}) cancel out, i.e. ${\cal E}_1^{\el,(1)}=0$, and we only have the vacuum-polarization contribution. Using the bound-state eigenfunction in Eq.~\eqref{psi0tilde}, we find the expression of the local electronic density matrix
\begin{eqnarray}
\b{n}_1^\el(x) &=& \frac{\kappa}{1+\lambda^2} \begin{pmatrix} 1 & -\i \lambda \sgn(x) \\ \i \lambda \sgn(x) & \lambda^2 \end{pmatrix} e^{-2\kappa |x|},
\label{n1elx}
\end{eqnarray}
where $\lambda$ was defined after Eq.~(\ref{1stboundarycond}). The electronic bound-state density is $n^{\el}(x)=\kappa e^{-2\kappa |x|}$ and the electronic current density vanishes, i.e. $j^\el(x)=0$.

This leads to the expression of the first-order direct and exchange Coulomb-type vacuum-polarization energy corrections to the bound-state energy
\begin{eqnarray}
{\cal E}_1^{\vp,(1),\text{DC}} 
\! &=& \! - \int_{-\infty}^{+\infty} \!\!\! \mathrm{d}x \int_{0}^{\infty} \!\!\frac{\mathrm d k}{\pi} \frac{\kappa^2 e^{-2\kappa |x|}}{k^2+\kappa^2} f_k(x),
\label{E1DC}
\end{eqnarray}
and
\begin{eqnarray}
{\cal E}_1^{\vp,(1),\text{XC}} 
\!\!&=&\!\frac{1}{1+\lambda^2} \!\! \int_{-\infty}^{+\infty} \!\!\! \mathrm{d}x \int_{0}^{\infty} \!\! \frac{\mathrm d k}{\pi} \frac{\kappa^2 e^{-2\kappa |x|} }{k^2+\kappa^2} 
\frac{\lambda^2-s_k^2}{1-s_k^2} f_k(x), \;\;\;\;
\label{E1XC}
\end{eqnarray}
and, similarly, to the first-order direct and exchange Breit-type vacuum-polarization energy corrections
\begin{eqnarray}
{\cal E}_1^{\vp,(1),\text{DB}} &=&0,
\label{E1DB}
\end{eqnarray}
and
\begin{eqnarray}
{\cal E}_1^{\vp,(1),\text{XB}} 
\!\! &=& \! \!\frac{-1}{1+\lambda^2} \!\! \int_{-\infty}^{+\infty} \!\!\!  \mathrm{d}x \int_{0}^{\infty} \!\! \frac{\mathrm d k}{\pi} \frac{\kappa^2 e^{-2\kappa |x|}}{k^2+\kappa^2}
\frac{1-\lambda^2 s_k^2}{1-s_k^2} f_k(x). \;\;\;\;\;
\label{E1XB}
\end{eqnarray}
Finally, after some simplifications, the total first-order vacuum-polarization energy correction can be put into the compact form
\begin{eqnarray}
{\cal E}_1^{\vp,(1)}\! &=&\! -\int_{-\infty}^{+\infty} \!\!\!\mathrm{d}x 
\int_{0}^{\infty} \!\!\frac{\mathrm d k}{\pi} \frac{\kappa^2 e^{-2\kappa |x|}}{k^2+\kappa^2} \Bigg(1 + \frac{\tilde{\varepsilon}_1 \varepsilon_k}{m^2c^4} \Bigg) f_k(x).
\end{eqnarray}

In the present 1D model, the direct contributions in Eqs.~(\ref{E1DC}) and~(\ref{E1DB}) and the exchange contributions in Eq.~(\ref{E1XC}) and~(\ref{E1XB}) are the equivalent of the direct and exchange contributions to the first-order QED ground-state energy correction in the 3D hydrogen-like atom with Coulomb potential~\cite{WicKro-PR-56,Moh-AP-74a,Moh-AP-74b}, which both contribute to the Lamb shift.

The different contributions to the first-order QED vacuum-polarization energy correction, as well as the total energy correction, are plotted in Fig.~\ref{fig:FirstOrderQEDCorrection} as a function of $Z$ for $c=137.036$ and as a function of $1/c$ for $Z=1$. As expected, as $Z$ increases or $1/c$ increases, the effect of relativity becomes stronger, and the different contributions increase in absolute value. The direct Coulomb-type correction is always negative and is the dominant contribution. The exchange Coulomb-type correction is always positive and the exchange Breit-type correction is always negative, these two contributions partially cancelling each other. In particular, in the low-relativistic regime ($Z \lesssim 40$ for $c=137.036$ or $1/c \lesssim 0.1$ for $Z=1$), the latter two contributions almost perfectly cancel each other. The total QED energy correction is always negative, leading thus to a stabilization of the bound state of the 1D hydrogen-like atom. This must be compared with standard QED in which the equivalent correction, after renormalization, on the ground-state energy of the 3D hydrogen-like atom contains a largely dominant positive exchange contribution (the ``self-energy'' contribution) and a much smaller negative direct contribution (the ``vacuum-polarization'' contribution), resulting in an overall destabilization of the ground state (see, e.g., Ref.~\onlinecite{EidGroShe-PR-01}). However, note that, just like in the 3D case, the QED energy correction in the present 1D model has the opposite sign than the leading relativistic energy correction in Eq.~(\ref{epsilon0tildelargec}), and thus tends to reduce the leading relativistic correction. Finally, using the Uehling-type approximation in Eq.~(\ref{nvp1xtext}), it can be inferred that the QED energy correction for the present 1D model starts at order $Z^2/c$ whereas for the 3D case the QED energy correction, after renormalization, starts at order $Z^4/c^3$ (see, e.g., Ref.~\onlinecite{SchPasPunBow-NPA-15}).

\section{Conclusion}
\label{sec:conclusion}

In this work, we have considered a 1D effective QED model of the relativistic hydrogen-like atom using delta-potential interactions. We have exposed the general exact theory, as well as the Hartree-Fock approximation. We have calculated the vacuum-polarization density at zeroth order in the two-particle interaction and the QED correction to the bound-state energy at first order in the two-particle interaction. The interest of this 1D toy model is that it shares the essential physical features of the 3D theory but eliminates some of the most serious technical difficulties coming from renormalization.

The next step will be to solve the present 1D effective QED model in a finite basis set with quantum-chemistry methods such as Hartree-Fock and configuration interaction. In particular, it will be interesting to understand how to efficiently represent the vacuum-polarization density in a finite basis set. This understanding should be very useful to reach the ultimate goal of having a fully-fledged quantum-chemistry implementation of 3D effective QED for atoms and molecules.

\section*{Acknowledgements}
The authors thank Trond Saue, Maen Salman, and Antoine Levitt for insightful discussions and useful comments on the manuscript.

\section*{Author Declarations}
The authors have no conflicts to disclose.

\section*{Data Availability}
The data that support the findings of this study are available from the corresponding author upon reasonable request.

\appendix
\section{Reduction of the Dirac Hamiltonian from 3D to 1D}
\label{app:3dto1d}

In 3D, we work in the Hilbert space $L^2(\mathbb{R}^3,\mathbb{C}) \otimes \mathbb{C}^4$ and the free-electron $4\times4$ Dirac Hamiltonian is
\begin{eqnarray}
\bm{{\cal D}}_0 = c \; (\vec{\bm{\alpha}} \cdot \vec{p}) +\bm{\beta} \; mc^{2},
\end{eqnarray}
where $\vec{p} = -i \vec{\nabla}$ is the momentum operator, and $\vec{\bm{\alpha}}$ and $\bm{\beta}$ are the $4 \times 4$ Dirac matrices
\begin{eqnarray}
\vec{\bm{\alpha}} = \left(\begin{array}{cc}
\b{0}_2&\vec{\bm{\sigma}}\\
\vec{\bm{\sigma}}&\b{0}_2\\
\end{array}\right)
~\text{and}~~
\bm{\beta} = \left(\begin{array}{cc}
\b{I}_{2}&\b{0}_2\\
\b{0}_2&-\b{I}_{2}\\
\end{array}\right),
\end{eqnarray}
where $\vec{\bm{\sigma}}=(\bm{\sigma}_1,\bm{\sigma}_2,\bm{\sigma}_3)$ is the 3-dimensional vector of the $2 \times 2$ Pauli matrices, and $\b{0}_2$ and $\b{I}_2$ are the $2 \times 2$ zero and identity matrices, respectively. The natural reduction of this Hamiltonian to the $x$-axis is the 1D free-electron $4\times4$ Dirac Hamiltonian
\begin{eqnarray}
\bm{{\cal D}}_{0,x} = c \; (\bm{\alpha}_1 \; p_x) +\bm{\beta} \; mc^{2}.
\end{eqnarray}

Using the unitary transformation
\begin{eqnarray}
\b{U} = \left(\begin{array}{cccc}
1 & 0 & 0 & 0\\
0 & 0 & 0 & 1\\
0 & 1 & 0 & 0\\
0 & 0 & 1 & 0\\
\end{array}\right),
\end{eqnarray}
the Hamiltonian $\bm{{\cal D}}_{0,x}$ can be transformed into the block-diagonal form
\begin{eqnarray}
\bm{{\cal D}}_{0,x}' = \b{U} \bm{{\cal D}}_{0,x}  \b{U}^{-1}= \left(\begin{array}{cc}
\b{D}_0 &\b{0}_2\\
\b{0}_2&\b{D}_0\\
\end{array}\right),
\label{3Dto1DdiracMatrix}
\end{eqnarray}
where $\b{D}_0 = c \; (\bm{\sigma}_1 \; p_x) +\bm{\sigma}_3 \; mc^{2}$ is the 1D free-electron $2\times2$ Dirac Hamiltonian introduced in Eq.~(\ref{freeDiracHam}). Correspondingly, the eigenstates of the block-diagonal Hamiltonian $\bm{{\cal D}}_{0,x}'$ can be chosen of the form
\begin{eqnarray}
\bm{\psi}_1' = \left(\begin{array}{c}
\psi^{\L}\\
\psi^{\S}\\
0\\
0\\
\end{array}\right) \;\;\text{and}\;\;
\bm{\psi}_2' = \left(\begin{array}{c}
 0\\
 0\\
 \psi^{\L}\\
 \psi^{\S}\\
\end{array}\right).
\label{psi1ppsi2p}
\end{eqnarray}
Thus, one can work simply with the $2\times2$ Dirac Hamiltonian $\b{D}_0$. The same result can also be obtained starting from the reduction of the Hamiltonian to the $y$-axis.

The eigenstates in Eq.~(\ref{psi1ppsi2p}) do not have definite spin, i.e. they are not eigenstate of the spin-projection operator $\bm{\Sigma}_3 = \bm{\sigma}_3 \oplus \bm{\sigma}_3$, where $\oplus$ designates the matrix direct sum. However, they have time-reversal symmetry. Indeed, the 1D Dirac Hamiltonian $\b{D}_0$ commutes with the 1D time-reversal operator~\cite{GuiMunPirSan-FP-19}, $\b{T}_\text{1D}= \bm{\sigma}_3 K_0$, where $K_0$ is the complex-conjugation operator. This implies that $\psi^{\L}$ and $\psi^{\S}$ can be chosen as real-valued and pure-imaginary functions, respectively, i.e. $\psi^{\L}=\psi^{\L*}$ and $\psi^{\S}=-\psi^{\S*}$. Imposing these constraints, we can show that $\bm{\psi}_1'$ and $\bm{\psi}_2'$ form a Kramers pair, i.e. they are connected by the 3D time-reversal operator~\cite{SauVis-INC-03,DyaFae-BOOK-07}, $\b{T}_\text{3D}= -\i \bm{\Sigma}_2 K_0$, where $\bm{\Sigma}_2 = \bm{\sigma}_2 \oplus \bm{\sigma}_2$. Indeed, applying the operator $\b{T}_\text{3D}$ in the new basis, we find
\begin{eqnarray}
\b{U} \b{T}_\text{3D} \b{U}^{-1} \bm{\psi}_1' = \left(\begin{array}{c}
0\\
0\\
\psi^{\L*}\\
-\psi^{\S*}\\
\end{array}\right)
= \left(\begin{array}{c}
0\\
0\\
\psi^{\L}\\
\psi^{\S}\\
\end{array}\right)
= \bm{\psi}_2',
\end{eqnarray}
and
\begin{eqnarray}
\b{U} \b{T}_\text{3D} \b{U}^{-1} \bm{\psi}_2' = \left(\begin{array}{c}
-\psi^{\L*}\\
\psi^{\S*}\\
0\\
0\\
\end{array}\right)
= \left(\begin{array}{c}
-\psi^{\L}\\
-\psi^{\S}\\
0\\
0\\
\end{array}\right)
= -\bm{\psi}_1'.
\end{eqnarray}

\section{Tensor product and partial trace}
\label{app:tensor}

We briefly review the tensor product (or Kronecker product) of vectors and matrices, and the concept of the partial trace.

Let us consider two vectors $\bm{\psi} \in\mathbb{C}^2$ and $\bm{\phi} \in\mathbb{C}^2$
\begin{eqnarray}
\bm{\psi} = \begin{pmatrix} \psi_1 \\ \psi_2 \end{pmatrix} ~~\text{and}~~\bm{\phi} = \begin{pmatrix} \phi_1 \\ \phi_2 \end{pmatrix}.
\end{eqnarray}
The tensor product of $\bm{\psi}$ and $\bm{\phi}$ is a vector $\bm{\Xi} \in \mathbb{C}^4$
\begin{eqnarray}
\bm{\Xi} = \bm{\psi} \otimes \bm{\phi} &=& 
\begin{pmatrix} \psi_1 \phi_1 \\ \psi_1 \phi_2 \\ \psi_2 \phi_1 \\ \psi_2 \phi_2 \end{pmatrix} = \begin{pmatrix} \Xi_{11} \\ \Xi_{12} \\ \Xi_{21} \\ \Xi_{22} \end{pmatrix},
\end{eqnarray}
where the elements $\Xi_{\rho \sigma} =  \psi_\rho \phi_\sigma$ are conveniently written with a composite index $\rho\sigma\equiv(\rho,\sigma) \in \{1,2\}^2$.
The tensor product of $\bm{\psi}^\dagger$ and $\bm{\phi}$ is a matrix $\b{M} \in \mathbb{C}^{2\times 2}$
\begin{eqnarray}
\b{M} = \bm{\psi}^\dagger \otimes \bm{\phi} &=& 
\begin{pmatrix}  \psi_1^* \phi_1 & \psi_2^* \phi_1 \\ \psi_1^* \phi_2 & \psi_2^* \phi_2 \end{pmatrix}
=\begin{pmatrix}  M_{1,1} & M_{1,2} \\ M_{2,1} & M_{2,2} \end{pmatrix}
,
\end{eqnarray}
with elements $M_{\rho, \sigma} =  \psi_\sigma^* \phi_\rho$.

Let us consider now two matrices $\b{A}\in \mathbb{C}^{2\times 2}$ and $\b{B}\in \mathbb{C}^{2\times 2}$
\begin{eqnarray}
\bf{A} = \begin{pmatrix} A_{1,1} & A_{1,2} \\ A_{2,1} & A_{2,2} \end{pmatrix}~~\text{and}~~ \bf{B} = \begin{pmatrix} B_{1,1} & B_{1,2} \\ B_{2,1} & B_{2,2}\end{pmatrix}.
\end{eqnarray}
The tensor product of $\b{A}$ and $\b{B}$ is a matrix $\b{C} \in \mathbb{C}^{4\times 4}$
\begin{eqnarray}
\b{C} = \bf{A} \otimes \bf{B} &=& 
\begin{pmatrix} 
A_{1,1} B_{1,1} & A_{1,1} B_{1,2} & A_{1,2} B_{1,1} & A_{1,2} B_{1,2}  \\
A_{1,1} B_{2,1} & A_{1,1} B_{2,2} & A_{1,2} B_{2,1} & A_{1,2} B_{2,2}  \\
A_{2,1} B_{1,1} & A_{2,1} B_{1,2} & A_{2,2} B_{1,1} & A_{2,2} B_{1,2}  \\
A_{2,1} B_{2,1} & A_{2,1} B_{2,2} & A_{2,2} B_{2,1} & A_{2,2} B_{2,2}  \\
\end{pmatrix}
\nonumber\\
&=&\begin{pmatrix} C_{11,11} & C_{11,12} & C_{11,21} & C_{11,22} \\ C_{12,11} & C_{12,12} & C_{12,21} & C_{12,22} \\ C_{21,11} & C_{21,12} & C_{21,21} & C_{21,22} \\ C_{22,11} & C_{22,12} & C_{22,21} & C_{22,22} \end{pmatrix},
\end{eqnarray}
with elements $C_{\rho \nu,\sigma \tau} = A_{\rho,\sigma}B_{\nu,\tau}$ written with composite indices $\rho\nu$ and $\sigma \tau$.

The (total) trace of $\b{C}$ is of course
\begin{eqnarray}
\Tr[\b{C}] = \sum_{\rho,\nu} C_{\rho \nu,\rho \nu} = \sum_{\rho,\nu} A_{\rho,\rho}B_{\nu,\nu} = \tr[ \b{A}] \; \tr[ \b{B}].
\end{eqnarray}
Due to the fact that $\b{C}$ is a tensor product of two matrices, we can also define a partial trace matrix $\Tr_1[\b{C}] \in \mathbb{C}^{2\times2}$ with respect to the first matrix $\b{A}$ (or the first ``particle''), with elements
\begin{eqnarray}
\left( \Tr_1[\b{C}] \right)_{\nu,\tau}= \sum_{\rho} C_{\rho \nu,\rho \tau} = \left( \sum_{\rho} A_{\rho,\rho} \right) B_{\nu,\tau},
\end{eqnarray}
i.e., $\Tr_1[\b{C}] = \tr[ \b{A}] \; \b{B}$. Similarly, we can define a partial trace matrix $\Tr_2[\b{C}] \in \mathbb{C}^{2\times2}$ with respect to the second matrix $\b{B}$ (or the second ``particle''), with elements
\begin{eqnarray}
\left( \Tr_2[\b{C}] \right)_{\rho,\sigma}= \sum_{\nu} C_{\rho \nu,\sigma \nu} = \left( \sum_{\nu} B_{\nu,\nu} \right) A_{\rho,\sigma},
\end{eqnarray}
i.e., $\Tr_2[\b{C}] = \tr[ \b{B}] \; \b{A}$.

\section{Commutator definition of the vacuum-polarization density}
\label{app:VPcommutator}

The vacuum-polarization density $n^{\vp}(x)$ in Eq.~(\ref{nvpx}) has been obtained with the normal-ordered definition of the density operator [see Eq.~\eqref{OPdensitymatrix}] 
\begin{eqnarray}
\hat{n}(x) = \tr\left( {\cal N}[\hat{\bm{\psi}}^\dagger(x)  \otimes \hat{\bm{\psi}}(x) ] \right).
\end{eqnarray}  
Another definition of the density operator commonly used in the literature (see, e.g., Refs.~\onlinecite{Sch-PR-49,WicKro-PR-56,NogBea-EL-86,HaiLewSerSol-PRA-07,HaiLewSol-CPAM-07,Tou-SPC-21,Sal-THESIS-22}) uses a commutator (c) instead of the normal ordering
\begin{eqnarray}
\hat{n}^\c(x) = \frac{1}{2} \tr\left( \left[\hat{\bm{\psi}}^\dagger(x),\hat{\bm{\psi}}(x) \right]_{\otimes} \right).
\end{eqnarray}  
where $[\hat{\bm{\psi}}^\dagger(x),\hat{\bm{\psi}}(x) ]_{\otimes} = \hat{\bm{\psi}}^\dagger(x) \otimes \hat{\bm{\psi}}(x) - \hat{\bm{\psi}}(x) \otimes \hat{\bm{\psi}}^\dagger(x)$ is the tensor-product commutator. With this definition, the corresponding vacuum-polarization density takes the form
\begin{eqnarray}
n^{\c,\vp}(x) = \frac{1}{2} \left(\sum_{p\in \NS} \tilde{\bm{\psi}}_p^\dagger(x) \tilde{\bm{\psi}}_p(x)- \sum_{p\in \PS} \tilde{\bm{\psi}}_p^\dagger(x) \tilde{\bm{\psi}}_p(x)\right).
\label{ncvpxdef}
\end{eqnarray} 
To calculate $n_1^{\c,\vp}(x)$, in the limits $L\to\infty$ and $\Lambda\to\infty$, we express the first sum over NS using the generalized negative-energy eigenfunctions in Eq.~\eqref{positronicWF}, and the second sum over PS using the bound-state eigenfunction in Eq.~\eqref{psi0tilde} and the generalized positive-energy eigenfunctions in Eq.~\eqref{electronicWF}, which leads to
\begin{eqnarray}
n^{\c,\vp}(x) = -\frac{\kappa e^{-2\kappa|x|}}{2} + \int_0^{\infty} \frac{\d k}{\pi} \frac{\kappa}{k^2+\kappa^2} \frac{\tilde{\varepsilon}_1}{\varepsilon_k} k \sin(2k|x|).
\label{ncvpx}
\end{eqnarray}
Using the relation
\begin{eqnarray}
\frac{\kappa e^{-2\kappa |x|}}{2} = \int_0^{\infty} \frac{\d k}{\pi} \frac{\kappa}{k^2+\kappa^2} \kappa \cos(2k|x|),
\end{eqnarray}
we see that $n^{\c,\vp}(x)$ is identical to $n^{\vp}(x)$ in Eq.~(\ref{nvpx})
\begin{eqnarray}
n^{\c,\vp}(x) = n^{\vp}(x).
\end{eqnarray}
However, we note that the expression of $n^{\c,\vp}(x)$ in Eq.~(\ref{ncvpx}) is subject to numerical instabilities for large $x$, contrary to the expression of $n^{\vp}(x)$ in Eq.~(\ref{nvpx}).

\section{Vacuum-polarization density from the Green function}
\label{app:green}

We derive here an alternative expression for the vacuum-polarization density $n^{\vp}(x)$ in Eq.~(\ref{nvpx}) based on the Green function.

The Green function (or resolvent) operator $\b{G}_0(\omega)=(\omega \b{I}_2 - \b{D}_0)^{-1}$ of the 1D free-electron Dirac Hamiltonian $\b{D}_0$ in Eq.~(\ref{freeDiracHam}) can be easily calculated in momentum space and Fourier-transformed back to real space (see, e.g., Refs.~\onlinecite{Lap-AJP-83,BenDab-LMP-94}), for $\omega \in \mathbb{C} \setminus \sigma(\b{D}_0)$ where $\sigma(\b{D}_0)$ is the spectrum of $\b{D}_0$,
\begin{eqnarray}
\b{G}_0(x,x';\omega) &=& -\frac{1}{2c} 
\left(\begin{array}{cc}
g(\omega) & \i \sgn(x-x')  \\
\i \sgn(x-x') & - g(-\omega)
\end{array}\right) e^{- \kappa(\omega) |x-x'|},
\nonumber\\
\label{G0xxp}
\end{eqnarray}
where $\kappa(\omega)= \sqrt{m^2 c^4-\omega^2}/c$ (with $\text{Re}[\kappa(\omega)] >0$) and $g(\omega) = \sqrt{(mc^2 + \omega)/(mc^2-\omega)}$. The Green function of the 1D hydrogen-like Dirac Hamiltonian $\b{D}$ in Eq.~(\ref{Dv}) satisfies the Dyson equation, for $\omega \in \mathbb{C} \setminus \sigma(\b{D})$,
\begin{eqnarray}
\b{G}(x,x';\omega) &=& \b{G}_0(x,x';\omega) 
\nonumber\\
&& \! + \! \int_{-\infty}^{+\infty}\!\!\! \d y \; \b{G}_0(x,y;\omega) \b{V}(y) \b{G}(y,x';\omega),
\label{}
\end{eqnarray}
where $\b{V}(y) = -Z \delta(y) \b{I}_2$, which gives
\begin{eqnarray}
\b{G}(x,x';\omega) = \b{G}_0(x,x';\omega) - Z \b{G}_0(x,0;\omega)  \b{G}(0,x';\omega).
\label{Gxxpomega}
\end{eqnarray}
In particular for $x=0$, we have
\begin{eqnarray}
\b{G}(0,x';\omega) = \b{G}_0(0,x';\omega) - Z \b{G}_0(0,0;\omega)  \b{G}(0,x';\omega),
\label{}
\end{eqnarray}
giving
\begin{eqnarray}
\b{G}(0,x';\omega) = [\b{I}_2 + Z \b{G}_0(0,0;\omega)]^{-1} \b{G}_0(0,x';\omega). 
\label{}
\end{eqnarray}
Inserting the last expression in Eq.~(\ref{Gxxpomega}), we obtain for the change of the Green function $\Delta \b{G}(x,x';\omega) = \b{G}(x,x';\omega) - \b{G}_0(x,x';\omega)$ 
\begin{eqnarray}
\Delta \b{G}(x,x';\omega) = \phantom{xxxxxxxxxxxxxxxxxxxxxxxxxxxxx}
\nonumber\\
- Z \b{G}_0(x,0;\omega) [\b{I}_2 + Z \b{G}_0(0,0;\omega)]^{-1} \b{G}_0(0,x';\omega).
\label{}
\end{eqnarray}
Finally, using the expression of $\b{G}_0(x,x';\omega)$ in Eq.~(\ref{G0xxp}), with the understanding that $\sgn(0)=0$, we obtain (see, e.g., Ref.~\onlinecite{BenDab-LMP-94})
\begin{eqnarray}
\Delta \b{G}(x,x';\omega) =  -\frac{Z}{4c^2} e^{- \kappa(\omega) (|x|+|x'|)} \phantom{xxxxxxxxxxxxxxx}
\nonumber\\
\times \left[ z_1(\omega) \b{G}_1(x,x';\omega)+ z_2(\omega) \b{G}_2(x,x';\omega) \right],
\label{DeltaGxxp}
\end{eqnarray}
with $z_1(\omega) = (1-\l \, g(\omega))^{-1}$, $z_2(\omega)=(1+\l \, g(-\omega))^{-1}$, $\l=Z/(2c)$, and $\b{G}_1(x,x';\omega)$ and $\b{G}_2(x,x';\omega)$ are the following matrices
\begin{eqnarray}
\b{G}_1(x,x';\omega) = \left(\begin{array}{cc}
g(\omega)^2 & -\i \sgn(x')g(\omega)  \\
\i \sgn(x) g(\omega) & \sgn(x) \sgn(x')
\end{array}\right),
\label{}
\end{eqnarray}
and
\begin{eqnarray}
\b{G}_2(x,x';\omega) = \left(\begin{array}{cc}
\sgn(x) \sgn(x') & -\i  \sgn(x)g(-\omega)  \\
\i \sgn(x') g(-\omega) & g(-\omega)^2
\end{array}\right).
\label{}
\end{eqnarray}

The vacuum-polarization density matrix (in the limits $L\to\infty$ and $\Lambda\to\infty$) is obtained by integrating $\Delta \b{G}(x,x';\omega)$ along the imaginary axis of frequency (see, e.g., Ref.~\onlinecite{HaiLewSerSol-PRA-07})
\begin{eqnarray}
\b{n}_1^\text{vp}(x,x') &=& \int_{-\infty}^{+\infty} \frac{\d u}{2\pi} \; \Delta \b{G}(x,x';\i u).
\label{n1vpfromG}
\end{eqnarray}
In particular, the vacuum-polarization density $n^\vp(x)$ is given by
\begin{eqnarray}
n^\text{vp}(x) &=& \int_{-\infty}^{+\infty} \frac{\d u}{2\pi} \; \text{tr}[ \Delta \b{G}(x,x;\i u) ],
\label{nvpfromG}
\end{eqnarray}
where, for $x\neq 0$,
\begin{eqnarray}
\text{tr}[\Delta \b{G}(x,x;\i u)]=-\frac{Z}{4c^2} e^{-2 \kappa(\i u) |x|} \phantom{xxxxxxxxxxxxxxx}
\nonumber\\
\times \left( z_1(\i u) \left( g(\i u)^2 +1\right) + z_2(\i u) \left( 1 + g(-\i u)^2 \right)\right).
\label{}
\end{eqnarray}
The integral in Eq.~(\ref{nvpfromG}) can be done numerically and perfectly matches the results from Eq.~(\ref{nvpx}).

We can also obtain the first-order vacuum-polarization density $n^{\text{vp},(1)}(x)$ with respect to $Z$, i.e. with respect to the external potential. It corresponds to setting $z_1(\omega)=1$ and $z_2(\omega)=1$, leading to
\begin{eqnarray}
\text{tr}[\Delta \b{G}^{(1)}(x,x;\i u)] =  -\frac{Z}{4c^2} e^{-2 \kappa(\i u) |x|} \left( 2+ g(\i u)^2 + g(-\i u)^2 \right),
\nonumber\\
\label{}
\end{eqnarray}
and
\begin{eqnarray}
n^{\text{vp},(1)}(x) &=& \int_{-\infty}^{+\infty} \frac{\d u}{2\pi} \; \text{tr}[ \Delta \b{G}^{(1)}(x,x;\i u) ] 
\nonumber\\
&=& -\frac{Z m^2 c^2}{\pi} \int_{0}^{\infty} \d u \; \frac{e^{-2 \sqrt{m^2c^4+u^2}|x|/c}}{m^2c^4+u^2}.
\label{}
\end{eqnarray}
Using the change of variables $t=\sqrt{1+(u/mc^2)^2}$, it can be expressed as
\begin{eqnarray}
n^{\text{vp},(1)}(x) &=&  -\frac{Z m}{\pi} \int_{1}^{\infty} \d t \; \frac{e^{-2 mc |x| t}}{t\sqrt{t^2-1}}.
\label{}
\end{eqnarray}
It is the equivalent for the present 1D model of the Uehling vacuum-polarization density (or potential) for the 3D hydrogen-like atom~\cite{Ueh-PR-35} (see also, e.g., Refs.~\onlinecite{GreRei-BOOK-09,IndMohSap-PRA-14}).

\section{Charge of the vacuum from the Green function}
\label{app:Nvac}

We calculate here the charge of the vacuum from Eqs.~(\ref{Nvac})-(\ref{Nvacp}) using the Green function.

We use the non-symmetry-adapted version of the generalized eigenfunctions of the free-particle Dirac Hamiltonian in Eqs.~(\ref{electronicFreeWF}) and~(\ref{positronicFreeWF}),
\begin{eqnarray}
\bm{\psi}_{+,k} (x)= B_k \left(\!\begin{array}{c}
1\\
s_k 
\end{array}\!\right) e^{\i k x} \;\text{and}\;
\bm{\psi}_{-,k} (x)= B_k \left(\!\begin{array}{c}
-s_k\\
1 
\end{array}\!\right) e^{\i k x},
\label{}
\end{eqnarray}
for $k\in \mathbb{R}$, and $B_k=\sqrt{(\varepsilon_k + mc^2)/(4\pi\varepsilon_k)}$. The projectors on the positive-energy and negative-energy eigenfunctions are then
\begin{eqnarray}
\b{P}_{+}^{0}(x',x) &=& \int_{-\infty}^{+\infty} \d k \; \bm{\psi}_{+,k}(x') \bm{\psi}_{+,k}^{\dagger}(x)
\nonumber\\
&=& \int_{-\infty}^{+\infty} \d k \; B_k^2\left(\!\begin{array}{cc}
1 & s_k\\
s_k  & s_k^2
\end{array}\!\right) e^{\i k (x'-x)},
\label{P+0}
\end{eqnarray}
and
\begin{eqnarray}
\b{P}_{-}^{0}(x',x) &=& \int_{-\infty}^{+\infty} \d k \; \bm{\psi}_{-,k}(x') \bm{\psi}_{-,k}^{\dagger}(x)
\nonumber\\
&=& \int_{-\infty}^{+\infty} \d k \; B_k^2\left(\!\begin{array}{cc}
s_k^2 & -s_k\\
-s_k  & 1 
\end{array}\!\right) e^{\i k (x'-x)}.
\label{P-0}
\end{eqnarray}
Inserting Eqs.~(\ref{P+0}) and~(\ref{n1vpfromG}) in Eq.~(\ref{Nvace}), and performing the inverse Fourier transformations over $x$ and $x'$, leads to the expression of the number of free electrons in the vacuum state
\begin{eqnarray}
N^\text{vac}_\text{e} &=& - Z \int_{-\infty}^{+\infty} \frac{\d u}{2\pi} \int_{-\infty}^{+\infty} \d k \; \frac{B_k^2}{(u^2+\varepsilon_k^2)^2} \phantom{xxxxxxxxx}
\nonumber\\
&&\;\;\;\;\;\;\;\;\;\times \left[ z_1(\i u) v_k (\i u) + z_2(\i u) w_k(\i u)\right],
\label{Nvaceint}
\end{eqnarray}
where $v_k (\omega) = [s_k(mc^2+\omega)-ck]^2$ and $w_k(\omega) = [(mc^2-\omega)+s_k ck]^2$. Similarly, inserting Eqs.~(\ref{P-0}) and~(\ref{n1vpfromG}) in Eq.~(\ref{Nvacp}) leads to the expression of the number of free positrons in the vacuum state
\begin{eqnarray}
N^\text{vac}_\text{p} &=& Z \int_{-\infty}^{+\infty} \frac{\d u}{2\pi} \int_{-\infty}^{+\infty} \d k \; \frac{B_k^2}{(u^2+\varepsilon_k^2)^2} \phantom{xxxxxxxxx}
\nonumber\\
&&\;\;\;\;\;\;\;\;\;\times \left[ z_1(\i u) w_k(-\i u) + z_2(\i u) v_k(-\i u)\right].
\label{Nvacpint}
\end{eqnarray}
Performing the integrals in Eqs.~(\ref{Nvaceint}) and~(\ref{Nvacpint}) numerically, we have checked that $N^\text{vac}_\text{e}=N^\text{vac}_\text{p}$ within the numerical precision.



\end{document}